\documentclass[pre,twocolumn,twoside,showpacs,superscriptaddress]{revtex4-1}
\usepackage{graphicx,amssymb,amsmath,epsf,bm}
\epsfclipon

\newcommand{\p}{\partial}
\newcommand{\ri}{\mathrm{i}}
\newcommand{\re}{\mathrm{e}}
\newcommand{\rd}{\mathrm{d}}

\newcommand{\prt}[2]{\frac{\partial #1}{\partial #2}}
\newcommand{\prts}[3]{\frac{\partial^{#3} #1}{\partial {#2}^{#3}}}
\renewcommand{\eqref}[1]{Eq.\ (\ref{#1})}
\newcommand{\eqsref}[1]{Eqs.\ (\ref{#1})}
\newcommand{\pref}[1]{(\ref{#1})}
\newcommand{\figref}[1]{Fig.\ \ref{#1}}

\newcommand{\const}{\text{const.}}  
\newcommand{\expct}[1]{\langle #1 \rangle}

\renewcommand{\[}{\left[}
\renewcommand{\]}{\right]}
\newcommand{\bnabla}{\bm{\nabla}}
\begin{document}

\title{Hyperbolic decoupling of tangent space and effective dimension of dissipative systems}

\author{Kazumasa A. Takeuchi}
\affiliation{Service de Physique de l'\'Etat Condens\'e, CEA-Saclay, F-91191 Gif-sur-Yvette, France}%
\affiliation{Department of Physics, The University of Tokyo,
 7-3-1 Hongo, Bunkyo-ku, Tokyo 113-0033, Japan}%
\author{Hong-liu Yang}
\affiliation{Institute of Physics, Chemnitz University of Technology, D-09107 Chemnitz, Germany}%
\author{Francesco Ginelli}
\affiliation{Physics Department, University of Florence, Via Sansone 1, I-50019 Sesto Fiorentino, Italy}%
\affiliation{Institut des Syst\`emes complexes de Paris \^Ile-de-France, 57-59 Rue Lhomond, F-75005 Paris, France}%
\affiliation{Istituto dei Sistemi Complessi (ISC), CNR, Via dei Taurini 19, I-00185 Roma, Italy}%
\author{G\"unter Radons}%
\affiliation{Institute of Physics, Chemnitz University of Technology, D-09107 Chemnitz, Germany}
\author{Hugues Chat\'e}
\affiliation{Service de Physique de l'\'Etat Condens\'e, CEA-Saclay, F-91191 Gif-sur-Yvette, France}%

\date{\today}

\begin{abstract}
We show, using covariant Lyapunov vectors, that
 the tangent space of spatially-extended dissipative systems
 is split into two hyperbolically decoupled subspaces:
 one comprising a finite number
 of frequently entangled ``physical'' modes,
 which carry the physically relevant information
 of the trajectory,
 and a residual set of strongly decaying ``spurious'' modes.
The decoupling of the physical and spurious subspaces
 is defined by the absence of tangencies between them
 and found to take place generally;
 we find evidence in partial differential equations
 in one and two spatial dimensions
 and even in lattices of coupled maps or oscillators.
We conjecture that the physical modes may constitute
 a local linear description of the inertial manifold
 at any point in the global attractor.
\end{abstract}


\maketitle

\section{Introduction}  \label{sec:introduction}

Nonlinear, dissipative, partial differential equations (PDEs)
 are frequently used for describing natural phenomena
 in all fields of physics and beyond.
They are particularly important
 for out-of-equilibrium problems such as pattern formation,
 turbulence, spatiotemporal chaos, etc.
 \cite{Cross_Hohenberg-RMP1993,Kuramoto-Book1984}.
Because of the nonlinearities involved,
 their generic solutions are mostly studied via numerical simulation.

A simple question arises then: can one faithfully integrate such PDEs,
 which are formally infinite-dimensional dynamical systems,
when numerical schemes obviously involve only a finite number 
of degrees of freedom? For the practitioner, the answer is yes, as one easily
observes that, increasing numerical resolution, the obtained (chaotic) 
solutions converge, in the sense that they exhibit the same dynamical or 
statistical properties. Mathematically speaking, the notion of
inertial manifold, on which all trajectories 
evolve after a short transient, is crucial in this context
\cite{Constantin_etal-Book1988,Foias_etal-JDiffEq1988,Robinson-Chaos1995}.
For generic parabolic PDEs such as the Kuramoto-Sivashinsky (KS) equation 
\cite{Kuramoto-Book1984}
 and the complex Ginzburg-Landau (CGL) equation \cite{Aranson_Kramer-RMP2002},
it is in fact proven
 that trajectories are first exponentially attracted
 to a finite-dimensional inertial manifold and eventually settle
 in a global attractor of finite Hausdorff dimension
 \cite{Foias_etal-JMathPuresAppl1988,Robinson-PhysLettA1994,Jolly_etal-AdvDiffEqs2000,Ghidaglia_Heron-PhysD1987,Doering_etal-Nonlinearity1988}.
The inertial manifold is a smooth object embedding the global attractor,
 which in principle reduces the evolution of trajectories
 to a finite set of ordinary differential equations.
However, this mathematical object remains largely formal,
for lack of a constructive way to determine
which modes actually constitute it.

In fact, even the dimension of the minimal inertial manifold stays beyond
 the reach of mathematical arguments.
So far, only upper bounds have been calculated, such as
 $L^{2.46}$ for the one-dimensional KS equation in a domain of size $L$
\cite{Robinson-PhysLettA1994,Jolly_etal-AdvDiffEqs2000}.
This is to be contrasted with the intuitive expectation that
 it should grow linearly with $L$,
 as suggested by the extensivity of chaos \cite{Ruelle-CMP1982}
 observed routinely in generic one-dimensional systems
 \cite{Manneville-Monograph1985,Tajima_Greenside-PRE2002,Keefe-PhysLettA1989,Livi_etal-JPhysA1986,Grassberger-PhysScr1989}
 including the KS equation
 \cite{Manneville-Monograph1985,Tajima_Greenside-PRE2002}.

In a different context, Cvitanovi\'c and coworkers 
have developed a constructive approach to unravel the ``skeleton''
of the chaotic dynamics of dissipative systems including PDEs 
\cite{Chaosbook,LanCvitanovic-PRE2008,Cvitanovic_etal-SIADS2010}:
by exploiting hierarchies of unstable periodic orbits, chaotic 
solutions are represented rather faithfully in simplified,
finite-dimensional spaces, and some quantitative properties can be 
systematically estimated. But it is not clear how the finiteness of the 
dimension of the inertial manifold is translated in this context, especially
since it is well known that there are infinitely-many unstable periodic orbits
involved, to various degrees, in the dynamics.

In the present paper, we show results that could constitute a first step
toward a constructive approach to the finite-dimensional 
inertial manifold of dissipative PDEs. 
Using Lyapunov analysis,
we study the tangent-space evolution of spatially-extended dynamical systems.
Since $N$-dimensional dynamical systems have $N$ Lyapunov exponents,
the numerical integration of a given PDE provides
 as many Lyapunov exponents as one likes
 by just increasing the spatial resolution.
Now, suppose that this PDE has an inertial manifold of finite dimension
and that we use a resolution with a larger number of degrees of freedom.
In a recent work \cite{Yang_etal-PRL2009}, 
 we showed that covariant Lyapunov vectors,
 which span the Oseledec subspaces \cite{Eckmann_Ruelle-RMP1985}
 and give the intrinsic directions of growth of perturbations
 for each Lyapunov exponent,
 exhibit a decomposition of tangent space into a fixed,
 finite number of ``physical'' modes 
 and a remaining set of ``spurious'' modes, whose number increases with 
increasing resolution. 
The excess spurious modes are associated to very negative Lyapunov exponents
 and ``hyperbolically'' decoupled from the physical modes.
Hence, perturbations along them quickly decay
 and do not spread to any physical mode.
Below, we pursue this approach, and provide
a precise characterization of the decoupling
 between the physical and spurious modes.
We show in particular that the decoupling takes place
 even between arbitrary combinations of physical or spurious modes,
 defining a finite-dimensional manifold in the tangent space,
 which should contain all the relevant information
 for the phase-space dynamics. 
We also provide a criterion giving an accurate estimate of 
 the dimension of the physical manifold from finite-time simulations.

As already mentioned above,
 the key quantities to access this hyperbolic decoupling
 are covariant Lyapunov vectors (CLVs).
At each point in phase space, they constitute the intrinsic directions of 
growth of perturbations associated with each Lyapunov exponent, in other words,
they span the Oseledec subspaces \cite{Eckmann_Ruelle-RMP1985}.
Therefore, their spatial structure is meaningful and their relative angles
rule the hyperbolicity properties of the dynamical system.
Note that they \textit{cannot} be replaced, even qualitatively,
 by the Gram-Schmidt vectors, which are by-products of the standard method
 \cite{Shimada_Nagashima-PTP1979,Benettin_etal-Meccanica1980}
 to compute Lyapunov exponents. It is only recently that CLVs have become
rather easy to compute numerically, notably thanks to the efficient algorithm
presented in \cite{Ginelli_etal-PRL2007}.
In the present study,
 we compute the Lyapunov exponents by the standard Gram-Schmidt method
 \cite{Shimada_Nagashima-PTP1979,Benettin_etal-Meccanica1980}
 and the associated CLVs by Ginelli's algorithm \cite{Ginelli_etal-PRL2007}.

The paper is organized as follows.
In Section\ \ref{sec:KS1d},
 we demonstrate our main results using the one-dimensional (1D) KS equation,
 a prototypical dissipative PDE showing spatiotemporal chaos.
We then apply the same approach to the 1D CGL equation
 in Sec.\ \ref{sec:CGL1d} to show how the hyperbolic decoupling
 reflects phase-space dynamics in different regimes of spatiotemporal chaos.
In particular we study what is called the phase turbulence 
regime \cite{Aranson_Kramer-RMP2002} of the CGL equation, 
in order to answer the long-standing question regarding whether this regime
can be fully described by ``phase modes''.
In Secs.\ \ref{sec:Lattice1d} and \ref{sec:KS2d}
 we study 1D lattice systems and the 2D KS equation,
 respectively, demonstrating the general existence of the physical manifold.
Section \ref{sec:discussion} contains a discussion and our conclusions.

\section{1D Kuramoto-Sivashinsky equation}  \label{sec:KS1d}

\subsection{Definition and numerical scheme}

We first focus on the 1D KS equation
 \cite{Cross_Hohenberg-RMP1993,Kuramoto-Book1984}:
\begin{equation}
 \prt{u}{t} = -\prts{u}{x}{2} - \prts{u}{x}{4} - u\prt{u}{x}, ~~~~ x \in [0,L],
 \label{eq:DefinitionKS1d}
\end{equation}
 with a real-valued field $u(x,t)$.
Boundary conditions are set to be periodic (PBC) $u(x,t) = u(x+L,t)$
 for most of the data below,
 but rigid boundary conditions (RBC) $u(0,t) = u(L,t) = 0$
 are also used for comparison.
The size is fixed at $L=96$ unless otherwise indicated,
 for which the 1D KS equation exhibits spatiotemporal chaos
 with both boundary conditions.
For numerical integration,
 we use the pseudospectral method
 with discrete Fourier or sine modes up to a cutoff wave number
 $k_{\rm cut}$.
The number of collocation points is chosen
 so that no aliasing may occur.
Integration is carried out with the operator-splitting method,
 which adopts the second-order Adams-Moulton method for the linear terms
 and Heun's method (two-stage second-order Runge-Kutta method)
 for the nonlinear term.
Most of the data presented in this section are obtained
 with PBC, $L=96$, and $k_{\rm cut} = 42 \cdot 2\pi/L$,
 i.e., with $43$ Fourier modes,
 from integration over a period of roughly $10^5$
 after a transient period about $10^4$ is discarded.

\subsection{Lyapunov exponents and Lyapunov vectors}

\begin{figure}[t]
 \includegraphics[width=\hsize,clip]{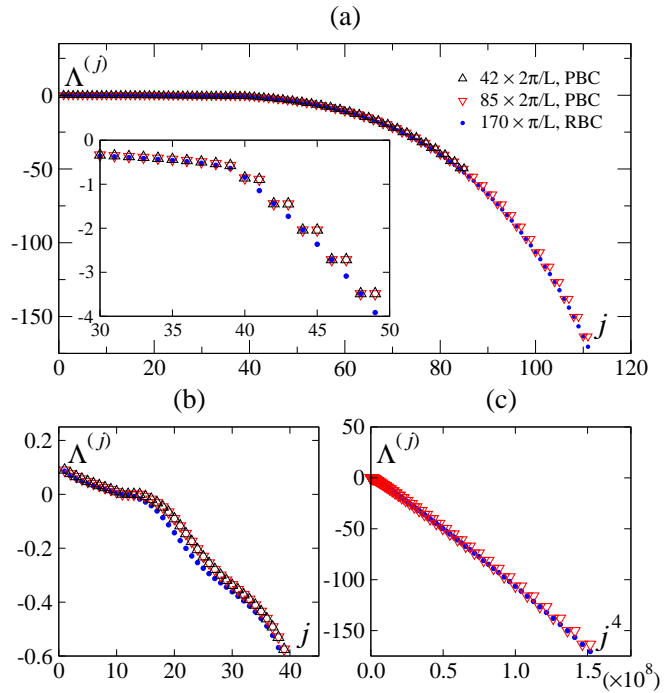}
 \caption{(color online). Spectrum of the Lyapunov exponents $\Lambda^{(j)}$ for the 1D KS equation with $L=96$. (a) Full spectra. Different symbols correspond to different cutoff wave numbers $k_{\rm cut}$ and different boundary conditions, either PBC or RBC, as indicated in the legend. Inset: Close-up around the threshold. (b) Close-up of (a) showing the positive end of the spectra. (c) The same Lyapunov spectra shown against $j^4$. The data for $k_{\rm cut} = 42 \cdot 2\pi/L$ with PBC (black upward triangles) are omitted for the sake of clarity. The value of the Kaplan-Yorke dimension is about $21.6$ for $k_{\rm cut} = 42 \cdot 2\pi/L$ and PBC.}
 \label{fig:KSLyapSpectra}
\end{figure}%

Figure \ref{fig:KSLyapSpectra}(a,b) shows the spectrum
 of the Lyapunov exponents $\Lambda^{(j)}$
 arranged in descending order for different spatial resolutions
 and boundary conditions.
The Lyapunov spectrum consists of two parts:
 first a smooth region of positive, zero, and negative exponents,
 and second a rather steep region of negative exponents
 arranged in steps of two for PBC.
The two regions are separated by an abrupt change in slope,
 accompanied by the formation of a stepwise structure for PBC
 (inset).
Remarkably, the spectra for different spatial resolutions
 overlap almost perfectly,
 with additional exponents due to the improved resolution
 simply accumulating at the negative end of the second region
 (upward and downward triangles).
Thus the threshold separating the two regions
 remains unchanged (here around $j=40$) upon increasing the resolution,
 though its exact index cannot be defined unambiguously
 only from the spectrum and will be given later on a firm basis.
Note also that in the second region
 the boundary conditions only change the multiplicity of modes:
 every other mode overlaps nicely for both types of boundary conditions.
All these observations lead to a speculation that
 the Lyapunov modes in the second region
 are ``spurious'' and dampen quickly, whereas all the physical properties
 of the dynamics are carried by the Lyapunov modes in the first region.
This intuition shall be substantiated in the following
 by studying the CLVs $\delta{}u^{(j)}(x,t)$
 associated with the Lyapunov exponents $\Lambda^{(j)}$.

\begin{figure}[t]
  \includegraphics[width=\hsize,clip]{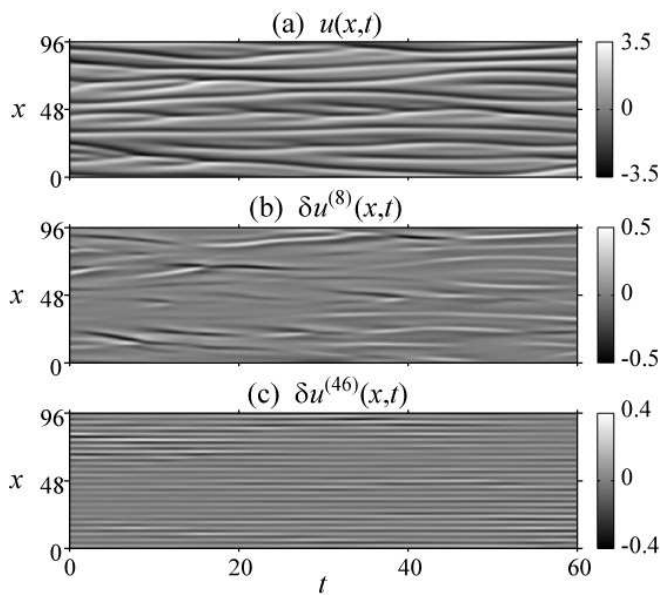}
  \caption{Typical spatiotemporal structure of the trajectory $u(x,t)$ (a) and CLVs $\delta u^{(j)}(x,t)$ (b,c) in the physical and the spurious regions, respectively, for the 1D KS equation ($L=96, k_{\rm cut} = 42 \cdot 2\pi/L$, PBC). The index of the CLV shown here is $j=8$ (b) and $j=46$ (c).}
  \label{fig:KSVectorStructures}
\end{figure}%

\begin{figure}[t]
  \includegraphics[width=\hsize,clip]{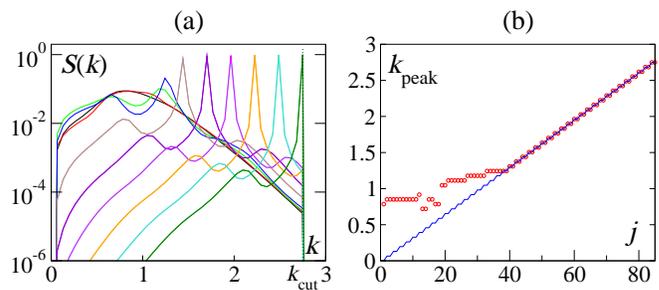}
  \caption{(color online). Spatial power spectrum of the CLVs for the 1D KS equation ($L=96, k_{\rm cut} = 42 \cdot 2\pi/L$, PBC). (a) Power spectrum of CLVs of index $j = 1, 16, 32, 38, 44, 52, 60, 68, 76$, and $84$ (from left to right at the peak positions). (b) Peak wave number $k_{\rm peak}$ of the power spectra (red circles) and $k=[j/2] \cdot 2\pi/L$ (blue line).}
  \label{fig:KSPowerSpectra}
\end{figure}

Figure \ref{fig:KSVectorStructures} shows typical spatiotemporal structures
 of CLVs in the physical and the spurious regions of the Lyapunov spectrum
 as well as that of the trajectory
 for the resolution $k_{\rm cut} = 42 \cdot 2\pi/L$ with PBC
 (corresponding to the black upward triangles in \figref{fig:KSLyapSpectra}).
At first glance, the CLVs for the physical and spurious Lyapunov modes
 have qualitatively different structures;
 whereas CLVs of the physical modes [\figref{fig:KSVectorStructures}(b)]
 evolve in a manner similar to the trajectory itself
 [\figref{fig:KSVectorStructures}(a)],
 those of the spurious modes are essentially sinusoidal
 apart from a slight modulation of the amplitude
 [\figref{fig:KSVectorStructures}(c)].
This can be seen more clearly in the spatial power spectra of the CLVs
 [\figref{fig:KSPowerSpectra}(a)].
In contrast to the smooth structure of the spectra
 found for the physical modes,
 those for the spurious modes are characterized by the existence
 of a sharp peak for the sinusoidal structure,
 whose wave number $k_{\rm peak}^{(j)}$ increases linearly
 with the index $j$ [\figref{fig:KSPowerSpectra}(b)].
For PBC, the two modes forming a step in the Lyapunov spectrum
 are found to have the same peak wave number and even the same power spectrum,
 but with an arbitrary phase shift in the vector structure.
This multiplicity of two does not exist with RBC,
 since the phase of the sinusoid is fixed at the boundary.
Moreover, the peak wave number $k_{\rm peak}^{(j)}$
 turns out to be simply the $j$th wave number,
 with the multiplicity taken into account,
 allowed in the given spatial geometry:
 $k_{\rm peak}^{(j)} = [j/2] \cdot 2\pi/L$ for PBC
 [\figref{fig:KSPowerSpectra}(b)],
 where $[x]$ is the integer part of $x$.
The power spectra of the spurious modes [\figref{fig:KSPowerSpectra}(a)]
 show that they are dominated by a sinusoidal structure
 at this trivial wave number $k_{\rm peak}^{(j)}$, and even more so as 
 $j$ is large.
Therefore, the value of their Lyapunov exponent is determined
 predominantly by the stabilizing linear term of the KS equation,
 i.e., the fourth-order derivative, and hence
 $\Lambda^{(j)} \approx -(k_{\rm peak}^{(j)})^4 \sim -j^4$
 for large enough $j$, as is indeed confirmed
 in \figref{fig:KSLyapSpectra}(c).

\subsection{Angles between covariant Lyapunov vectors}

The sinusoidal structure of the Lyapunov vectors of spurious modes suggests
that they are nearly orthogonal to each other.
Unlike Gram-Schmidt vectors, CLVs allow us to check this directly
 through the angle $\theta^{(i,j)}$ between vectors
 defined by the inner product
\begin{equation}
 \cos \theta^{(i,j)}(t)
 \equiv \int_0^L \delta u^{(i)}(x,t) \delta u^{(j)}(x,t) \rd x,  \label{eq:DefinitionAngle}
\end{equation}
 with $L^2$-normalized vectors
 $\int_0^L \delta u^{(j)}(x,t)^2 \rd x = 1$.
Since the angle $\theta$ fluctuates in time,
 one needs to study its distribution function $\rho_{\rm v}(\theta)$.

\begin{figure}[t]
  \includegraphics[width=\hsize,clip]{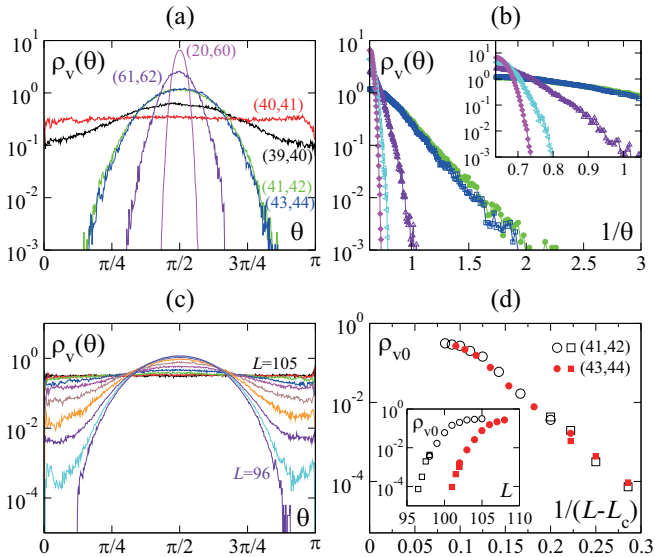}
  \caption{(color online). Distributions $\rho_{\rm v}(\theta)$ of the angles $\theta$ between pairs of CLVs for the 1D KS equation ($k_{\rm cut} = 42 \cdot 2\pi/L$, PBC). (a) $\rho_{\rm v}(\theta)$ versus $\theta$ for $L=96$. The indices of the pairs are indicated on the figure. (b) $\rho_{\rm v}(\theta)$ versus $1/\theta$ for $L=96$ and for pairs (41,42), (43,44), (61,62), (81,82), (20,60) from right to left. The ordinate is averaged over both sides of the distribution. Inset: close-up of the distributions. (c) Angle distributions $\rho_{\rm v}(\theta)$ for the pair (41,42) with varying system size $L = 96, 97, \dots, 105$ (from inside to outside in the figure). The distribution is obtained from shorter simulations for $L>99$, recorded over time $10^5$. (d) Probability density $\rho_{{\rm v}0}$ of the distribution $\rho_{\rm v}(\theta)$ at $\theta = 0$ and $\pi$, as a function of $L$ (inset) and of $1/(L-L_{\rm c})$ with a given $L_{\rm c}$ (main panel). Pairs of the vectors are fixed at $(41,42)$ (black open symbols) and $(43,44)$ (red full symbols). The value of $L_{\rm c}$ is $L_{\rm c} = 93.0$ for the pair $(41,42)$ and $L_{\rm c} = 97.5$ for $(43,44)$. The circles and squares correspond to the recording time $10^5$ and $10^6$, respectively.}
  \label{fig:KSAngleDist}
\end{figure}%

The result is shown in \figref{fig:KSAngleDist}(a) for PBC.
Indeed, the angle distributions $\rho_{\rm v}(\theta)$
 between any pairs of CLVs of index $j \geq 42$ are peaked at $\pi/2$
 and drop rapidly near $0$ and $\pi$ (green, blue, and indigo curves).
This is also true if one of the vectors is taken in the region
 $j \leq 41$ (purple curve), but when both vectors are taken from this region
 (black and red curves),
 the angle distribution spans the whole $[0,\pi]$ interval.
The change in the distribution is found to be quite sharp in this example;
 notice that the indices are varied only one by one
 for neighboring pairs in \figref{fig:KSAngleDist}(a)
 (the pairs in the same step, e.g., $(42,43)$, are skipped).
The transition between physical and spurious modes
 can thus be characterized by the existence or the absence
 of vector tangencies, arbitrarily close to the phase-space trajectory.
An angle distribution bounded away from $0$ and $\pi$
 marks the absence of such tangencies,
 while a finite weight at $0$ or $\pi$ indicates that
 the given pair of vectors has a finite probability
 of forming any arbitrarily small angle along the phase-space trajectory.
Note that while two different, non-degenerate CLVs,
 will never be exactly coincident, thus tangent,
 at any point of the phase-space trajectory,
 they can get arbitrarily close to each other
 if the angle distribution has a finite value at $0$ or $\pi$.
In the following,
 we will use the term tangency to describe this latter occurrence.
With this in mind, the results in \figref{fig:KSAngleDist}(a)
 show that the physical modes exhibit such tangencies rather frequently.
In contrast,
 the spurious modes,
 in addition to the negative values of their Lyapunov exponent,
 do not have any tangencies
 with other spurious modes (except the partner in the same step)
 nor with physical modes.
In this sense, they are hyperbolically isolated
 from any other modes.

A closer look reveals that,
 while at the threshold index
 the tail of the angle distribution decreases as
 $\rho_{\rm v}(\theta) \sim \exp(-\const/\theta)$  
 with decreasing $\theta$ within our numerical precision,
 showing an essential singularity at $\theta = 0$ and $\pi$,
 it decays even more rapidly for larger indices
 [\figref{fig:KSAngleDist}(b)].
This qualitative change of behavior provides, in our opinion, 
a more accurate way of defining the threshold.
Nevertheless, care should be taken
 when arguing about the existence or the absence of tangencies
 from finite-time simulations.
To determine the exact threshold,
 one needs to perform a careful asymptotic analysis
 on the frequency of tangencies.
This can be achieved by studying the system size dependence
 of the angle distribution $\rho_{\rm v}(\theta)$.
Figure \ref{fig:KSAngleDist}(c) shows
 how  $\rho_{\rm v}(\theta)$ changes for a given pair of CLVs,
 the pair (41,42) here, with varying system size $L$.
The value of $\rho_{\rm v}(\theta)$ at the tail,
 denoted by $\rho_{{\rm v}0}$, decreases with decreasing $L$,
 and from a certain system size the tail disappears
 [inset of \figref{fig:KSAngleDist}(d)].
A singularity analysis shows that it decays
 as $\rho_{{\rm v}0} \sim \exp[-\const/(L-L_{\rm c})]$,
 providing a threshold
 $L_{\rm c}=93$ for the pair (41,42) and $L_{\rm c}=97.5$ for (43,44)
 [main panel of \figref{fig:KSAngleDist}(d)].
This in turn
 allows us to determine the exact number of the threshold index,
 or the number of the physical modes, at $N_{\rm ph}=43$ for $L=96$.
The critical decay of the angle distribution observed for a fixed size $L$,
 $\rho_{\rm v}(\theta) \sim \exp(-\const/\theta)$,
 remains up to $\theta\to 0$, strictly,
 only at $L=L_{\rm c}$ for a given pair of indices;
 otherwise it should be eventually replaced
 by a faster decay or convergence to a finite value.

Be aware that the threshold $N_{\rm ph}$ determined here does \textit{not}
 coincide with the index at which the stepwise structure
 of the Lyapunov spectrum is formed for PBC
 [$j=40$ here; see inset of \figref{fig:KSLyapSpectra}(a)].
Indeed, a closer scrutiny of the exponents reveals that
 the first few steps are actually slightly inclined,
 and hence no threshold can be unambiguously defined
 from them. 
The exponents alone can only provide a good guess
 of the location of the threshold;
 instead, it must be determined from the hyperbolicity properties
 as evidenced in this section.

\subsection{Angles between subspaces}

In the previous section we showed the hyperbolic decoupling
 of all the spurious modes from any physical mode.
This, however, does not necessarily imply
 that they are also decoupled from the manifold
 spanned by the physical modes,
 as properly pointed out by Kuptsov and Kuznetsov
 \cite{Kuptsov_Kuznetsov-PRE2009},
 since it does not exclude the possibility of tangencies
 between a linear combination of physical modes and of spurious ones.
Here, following the idea in Ref.\ \cite{Kuptsov_Kuznetsov-PRE2009},
 we clarify this issue by studying
 the smallest possible angle formed by any linear combination
 of physical modes
 and any linear combination of spurious modes.
It is equivalent to the minimum angle between the subspaces spanned by them.

The angle between subspaces can be computed as follows:
Given the matrices $Q_{\rm ph}$ and $Q_{\rm sp}$
 defining the orthogonal bases of the two subspaces,
 which can be easily obtained by the QR decomposition
 of the arrays of the corresponding CLVs,
 the minimum angle between the two subspaces can be obtained
 simply from the singular value decomposition of the matrix
 $Q_{\rm ph}^{\rm T} Q_{\rm sp}$
 \cite{Bjorck_Golub-MathComput1973,Knyazev_Argentati-SIAMJSciComput2002},
 as
\begin{equation}
 \cos \phi = \sigma_1(Q_{\rm ph}^{\rm T} Q_{\rm sp}),  \label{eq:DefinitionSubspaceAngle}
\end{equation}
 where $\sigma_1$ is the largest singular value
 and the superscript $^{\rm T}$ denotes the transpose \cite{Note1}.

\begin{figure}[t]
  \includegraphics[width=\hsize,clip]{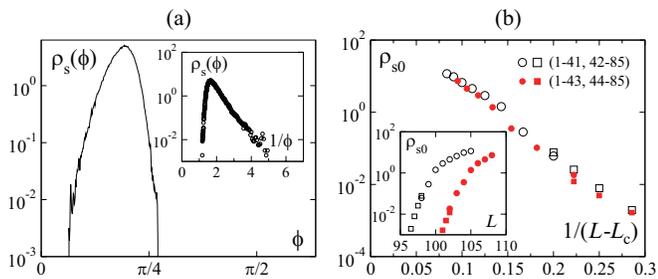}
  \caption{(color online). Distribution $\rho_{\rm s}(\phi)$ of the angle $\phi$ between the physical and spurious subspaces for the 1D KS equation ($k_{\rm cut} = 42 \cdot 2\pi/L$, PBC). (a) $\rho_{\rm s}(\phi)$ versus $\phi$ for $L=96$. The physical subspace is spanned by CLVs of index $1$ to $N_{\rm ph} = 43$, while the rest spans the spurious subspace. Note that $\phi$ is by definition equal to or smaller than $\pi/2$, in contrast to $\theta$ in \figref{fig:KSAngleDist}. Inset: the same data shown against $1/\phi$. (b) Probability density $\rho_{{\rm s}0}$ of the distribution $\rho_{\rm s}(\theta)$ at $\phi=0$ for varying system size $L$. It is measured between two subspaces spanned by CLVs of indices given in the legend, and shown against $L$ (inset) and $1/(L-L_{\rm c})$ (main panel) with the same $L_{\rm c}$ as in \figref{fig:KSAngleDist}(d). The circles and squares correspond to the recording time $10^5$ and $10^6$, respectively.}
  \label{fig:KSSubspaceAngleDist}
\end{figure}%

Figure \ref{fig:KSSubspaceAngleDist} shows the result,
 which does not change the view presented above.
The angle distribution $\rho_{\rm s}(\phi)$ is bounded away from zero
 [\figref{fig:KSSubspaceAngleDist}(a)],
 so that no tangency exists even between the two subspaces.
The tail of the distribution is governed by the pairs of Lyapunov modes
 of closest indices from both groups,
 and hence we find a similar critical decay
 $\rho_{\rm s}(\phi) \sim \exp(-\const/\phi)$
 within the observed time window (inset).
The system size dependence is also studied in the same manner
 [\figref{fig:KSSubspaceAngleDist}(b)],
 providing the same value of $L_{\rm c}$ as from the vector angle distribution
 $\rho_{\rm v}(\theta)$ shown in \figref{fig:KSAngleDist}(d).
This implies that the subspace spanned by the $N_{\rm ph}$ physical modes
 is hyperbolically decoupled from the subspace of the remaining spurious modes
 or from any other single spurious mode.
In strong contrast, inside the manifold of the physical Lyapunov modes,
 these modes are highly entangled with frequent tangencies
 as we have found in \figref{fig:KSAngleDist}(a).

\subsection{Domination of Oseledec splitting}

The absence of tangencies for the spurious modes can also be confirmed
 from another viewpoint,
 specifically, in terms of the domination of the Oseledec splitting (DOS)
 \cite{Pugh_etal-BullAmMathSoc2004,Bochi_Viana-AnnMath2005}.
Roughly, DOS refers to the dynamical isolation of the Oseledec subspaces
 from each other due to the strict ordering
 of the local expansion rates.
Let $\lambda_\tau^{(j)}(t)$ be the finite-time Lyapunov exponent
 obtained by averaging the local expansion rate of the $j$th CLV
 from time $t$ to $t+\tau$, i.e.,
\begin{equation}
 \lambda_\tau^{(j)}(t) \equiv \frac{1}{\tau} \log ||J_\tau(t) \delta u^{(j)}(x,t)||,
\end{equation}
 where $J_\tau(t)$ is the Jacobian operator for the evolution
 of the given dynamical system over time $\tau$
 and the CLVs are normalized as
 $||\delta u^{(j)}(x,t)||^2 \equiv \int_0^L \delta u^{(j)}(x,t)^2 \rd x = 1$.
Then, the splitting of a pair of one-dimensional subspaces,
 or CLVs, of indices $j_1$ and $j_2 (> j_1)$
 is said to be dominated
 if $\lambda_\tau^{(j_1)}(t) > \lambda_\tau^{(j_2)}(t)$ holds
 for all $t$ with $\tau$ larger than a finite $\tau_0$.
It has been mathematically proven that DOS implies the absence of tangency
 between the corresponding CLVs
 \cite{Pugh_etal-BullAmMathSoc2004,Bochi_Viana-AnnMath2005}.
To quantify DOS, we define, following \cite{Yang_Radons-PRL2008},
\begin{equation}
 \Delta\lambda_\tau^{(j_1,j_2)}(t) \equiv \lambda_\tau^{(j_1)}(t) - \lambda_\tau^{(j_2)}(t) \;.  \label{eq:DefinitionDeltaLambda}
\end{equation}
We then measure the time fraction of the DOS violation
\begin{equation}
 \nu^{(j_1,j_2)}_\tau = \langle \Theta( \Delta\lambda_\tau^{(j_1,j_2)}(t) ) \rangle,  \label{eq:DefinitionNu}
\end{equation}
 where $\Theta(z)$ is the step function
 and the brackets denote the time average.
Note that one needs to compute
 the finite-time Lyapunov exponents $\lambda_\tau^{(j)}(t)$
 from the CLVs,
 \textit{not} from the standard method based on the QR decomposition
 \cite{Shimada_Nagashima-PTP1979,Benettin_etal-Meccanica1980},
 because they indicate different local expansion rates
 despite the same values of the long-time average.
The right ones reflecting the physical properties
 of the tangent space, such as DOS, are those from the CLVs
 \cite{Pugh_etal-BullAmMathSoc2004,Bochi_Viana-AnnMath2005}.

\begin{figure}[t]
  \includegraphics[width=\hsize,clip]{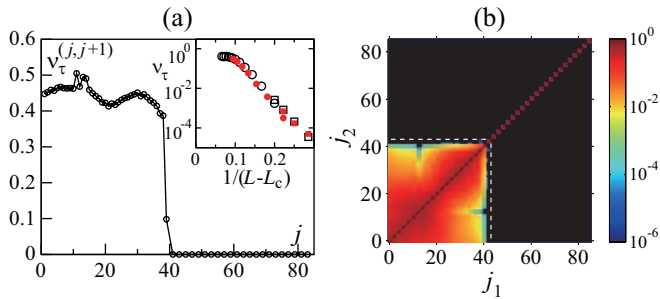}
  \caption{(color online). Violation of DOS in the KS equation ($L=96, k_{\rm cut} = 42 \cdot 2\pi/L$, PBC). (a) Time fraction of the DOS violation, $\nu^{(j,j+1)}_\tau$, for pairs of the neighboring modes with $\tau = 0.2$. The pairs within the same step are omitted. Inset: $\nu^{(j,j+1)}_{\tau=0.2}$ with $j=41$ (black open symbols) and $j=43$ (red full symbols), shown against $1/(L-L_{\rm c})$ with the same $L_c$ as in \figref{fig:KSAngleDist}(d). The circles and squares correspond to the recording time $10^5$ and $10^6$, respectively. (b) $\nu^{(j_1,j_2)}_{\tau=0.2}$ for arbitrary pairs. The black color indicates $\nu^{(j_1,j_2)}_\tau = 0$, i.e., hyperbolically isolated pairs. The white dashed lines indicate the threshold $N_{\rm ph}=43$ between the physical and spurious modes.}
  \label{fig:KSDosViolation}
\end{figure}%

The result is shown in \figref{fig:KSDosViolation}.
The DOS violation fraction $\nu^{(j,j+1)}_\tau$ for the pairs
 of the neighboring exponents with $\tau = 0.2$
 drops sharply near the threshold and becomes strictly zero
 for $j \geq N_{\rm ph}=43$ [\figref{fig:KSDosViolation}(a)]
 except the pairs in the same step.
This split is found not only for the pairs of the neighbors;
 in fact $\nu^{(j_1,j_2)}_\tau$ stays zero for any pair
 that includes a spurious mode [\figref{fig:KSDosViolation}(b)],
 as expected from the fact that all the spurious modes are
 hyperbolically decoupled from the physical modes.
The exact number of the physical modes $N_{\rm ph}$ can also be checked
 from the system size dependence of $\nu^{(j,j+1)}_\tau$
  [inset of \figref{fig:KSDosViolation}(a)]
 in the same way as for the angle distributions
 of the vectors or the subspaces.
All this confirms the absence of tangencies of the spurious modes,
 which has already been seen from the angle distributions,
 indicating the same number of the physical modes.

\subsection{What does the absence of tangencies imply?}

We now explore the implication of the absence of tangencies
 for the spurious modes.
Suppose that we add an infinitesimal perturbation along an arbitrary CLV
 to the dynamics.
If this is a spurious mode, the perturbation decays exponentially to zero
 as indicated by their negative Lyapunov exponents.
Further, the absence of tangencies implies that
 this does not induce any perturbation along the directions spanned
 by the other Lyapunov modes.
In contrast, perturbations along the physical modes will
 propagate to other physical modes through the tangencies between them,
 that is, whenever two modes get arbitrarily close to each other,
 and eventually induce activity in all the physical modes.
This is true even if the initial perturbation was made
 along the physical modes with negative exponents,
 since they are directly or indirectly
 connected to those with positive exponents,
 where the propagated perturbation exponentially grows
 to affect the phase-space dynamics considerably.
The situation does not change if we initially add a perturbation
 along a linear combination of physical or spurious CLVs:
 since the physical and spurious subspaces themselves are hyperbolically
 isolated (\figref{fig:KSSubspaceAngleDist}),
 activities in the spurious modes decay altogether
 without propagating to one or a set of physical modes, and vice versa.
In this sense, the dynamics corresponding to the physical modes
 is highly entangled, but completely decoupled from
 the decaying dynamics of the spurious modes.

Moreover, we find that the number of the physical modes $N_{\rm ph}$
 does not depend on the spatial resolution, provided it is high enough.
This implies that even in the high resolution limit, or in the original PDE,
 the number of the physical modes remains the same, 
 while infinitely many spurious modes can appear by increasing the resolution.
Therefore, the physical modes are probably associated with
 the finite number of the intrinsic degrees of freedom
 describing the dynamics of the PDE.
This leads to the conjecture
 that the physical modes may span the local linear approximation
 of the (minimal) inertial manifold.
The number of the physical modes $N_{\rm ph}$ would then indicate
 the inertial manifold dimension,
 which is the smallest possible number of degrees of freedom
 to describe the dynamics of the given system.

\subsection{Extensivity}

\begin{figure}[t]
  \includegraphics[clip]{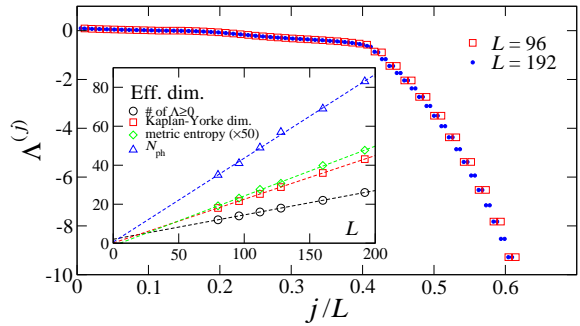}
  \caption{(color online). Extensivity of the Lyapunov spectrum (main panel) and effective dimensions (inset) of the KS equation (with PBC). Shown in the inset are the number of the non-negative Lyapunov exponents (black circles), the Kaplan-Yorke dimension (red squares), the metric entropy (green diamonds, multiplied by $50$), and the number of the physical modes $N_{\rm ph}$ (blue triangles). $N_{\rm ph}$ is estimated here from finite-time simulations of given sizes, so that it could be slightly underestimated. The dashed line indicates linear regression for each quantity.}
  \label{fig:KSExtensivity}
\end{figure}%

The physical modes being the active degrees of freedom of the system,
 one expects that their number $N_{\rm ph}$
 scales linearly with the system size $L$ in systems known otherwise to 
show extensive chaos.
This is indeed the case:
 in \figref{fig:KSExtensivity}
 the Lyapunov spectra with their index rescaled by $L$ collapse
 both in the physical and in the spurious regions with the 
 stepwise structure retained.
As suggested from this collapse, the number of physical modes
 is found to increase linearly with $L$, i.e., it is an extensive quantity,
 just like the other common quantities indicating effective dimensions
 of a dynamical system \cite{Eckmann_Ruelle-RMP1985}
 such as the number of the non-negative Lyapunov exponents and
 the Kaplan-Yorke dimension
 (inset of \figref{fig:KSExtensivity}).
Note that the physical manifold dimension $N_{\rm ph}$ (blue triangles)
 is the largest,
 in particular, larger than the Kaplan-Yorke dimension (red squares).
This corroborates the speculation made above
 that the physical manifold may provide
 a tangent-space representation of the inertial manifold,
 which is a smooth manifold embedding the global attractor.
Though, clearly, some mathematical rigor is needed here,
our result may provide an estimate of the KS inertial manifold dimension $D$
 at
\begin{equation}
 D = N_{\rm ph} \approx 0.43 \times L,  \label{eq:KSNumPhys}
\end{equation}
 which drastically lowers the existing mathematical upper bounds
 $D \leq \const \times L^{2.46}$
 \cite{Robinson-PhysLettA1994,Jolly_etal-AdvDiffEqs2000}.
Given the length scale $2\pi\sqrt{2}$ associated with
 the most unstable wave number of the KS equation,
 which can be taken as the size of the ``building blocks''
 of spatiotemporal chaos in this system,
 the above estimate implies that each building block contains roughly
 $2\pi\sqrt{2} \times 0.43 \approx 3.8$ degrees of freedom.

In fact, the estimate in \eqref{eq:KSNumPhys} is obtained
 from the angle distributions and the DOS violation of finite-time simulations,
 so that the prefactor in \eqref{eq:KSNumPhys} may be slightly underestimated.
If we use, instead, the fact that $N_{\rm ph}$ increases exactly by two
 between two threshold sizes $L_{\rm c} = 93.0$ and $97.5$,
 we obtain $N_{\rm ph} \approx 0.44 \times L$,
 or $4.0$ physical modes
 per building block of length $2\pi\sqrt{2}$.
This estimate is remarkably close to
 the result of an argument by Sasa \cite{Sasa-PC1} showing that
 solutions of the 1D KS equation can be reconstructed,
 at least qualitatively,
 with four modes per building block.

\section{1D Complex Ginzburg-Landau Equation}  \label{sec:CGL1d}

\subsection{Definition and numerical scheme}

To elucidate the generality of our finding, we now consider
 the CGL equation.
It describes oscillatory modulations in spatially extended continuous media
 at the most basic level,
 but the genericity of the phenomena described by the CGL equation is known
to be in fact much larger
 \cite{Cross_Hohenberg-RMP1993,Aranson_Kramer-RMP2002}.
In one spatial dimension, it governs a complex field
 $W(x,t)$ with $x \in [0,L]$ according to
\begin{equation}
 \prt{W}{t} = W - (1+\ri\beta)|W|^2 W + (1+\ri\alpha)\prts{W}{x}{2},
 \label{eq:DefinitionCGL}
\end{equation}
 for which the phase $\Phi(x,t)$ associated with the isochrone
 \cite{Kuramoto-Book1984}
 of the local oscillation is given by
\begin{equation}
 \Phi(x,t) = \arg W(x,t) - \beta \ln|W(x,t)|.  \label{eq:DefinitionCGLPhase}
\end{equation}
We consider here only PBC, $W(x+L,t) = W(x,t)$.
Numerical integration is performed similarly to the KS equation,
 with the pseudospectral method and the operator-splitting method
 implemented by the second-order Adams-Moulton and Heun's methods.

\begin{figure}[t]
  \includegraphics[width=\hsize,clip]{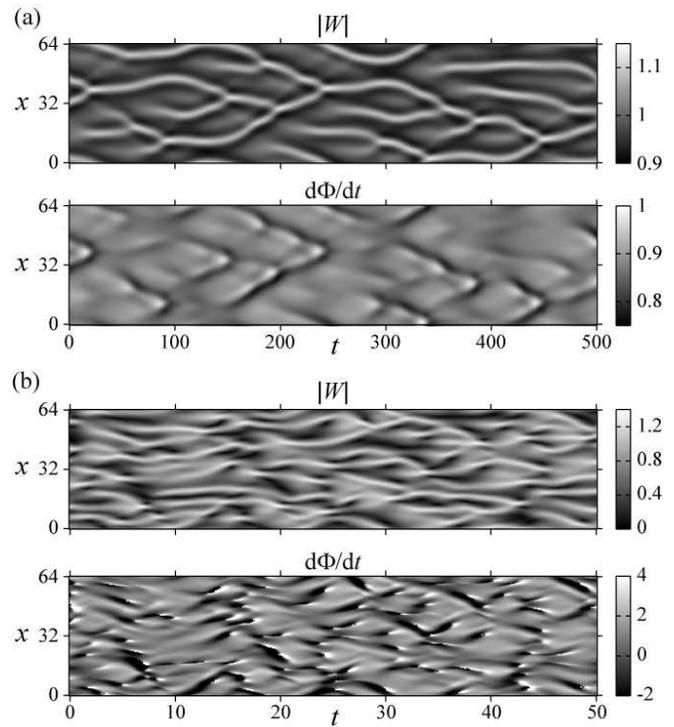}
  \caption{Spatiotemporal dynamics of the phase turbulence at $\alpha = 2.0$ and $\beta = -0.91$ (a) and the amplitude turbulence at $\alpha = 2.5$ and $\beta = -2.0$ (b) in the CGL equation with PBC. The system size is $L = 64$ for both cases. The gray scale covers the whole range of the values taken by $|W|$ and $\rd\Phi/\rd t$ except for the phase velocity $\rd\Phi/\rd t$ in the amplitude turbulence (b) showing frequent phase slips (bright and dark spots). Notice the different time scales between (a) and (b).}
  \label{fig:CGLSnapshots}
\end{figure}%

We study here two representative regimes
 of the spatiotemporal chaos of the CGL equation,
 namely the amplitude turbulence regime and the phase turbulence regime
 \cite{Aranson_Kramer-RMP2002,Shraiman_etal-PhysD1992,Chate-Nonlinearity1994}.
Both regimes correspond to the situation where
 plane-wave solutions are rendered unstable by the Benjamin-Feir instability.
When varying the parameter values
 and crossing the Benjamin-Feir line at $\alpha\beta=-1$,
 one first encounters the phase turbulence regime;
 the phase field $\Phi(x,t)$ controls the spatiotemporal chaos,
 while the amplitude $|W(x,t)|$ remains rather quiescent,
 staying nonzero everywhere
 [\figref{fig:CGLSnapshots}(a)].
Beyond this regime, amplitude turbulence takes place,
 in which defects associated with null amplitude and discontinuous change
 in the phase, in other words phase slips,
 are constantly created and annihilated [\figref{fig:CGLSnapshots}(b)].
The observed dynamics is chaotic both in amplitude and in phase
 for the amplitude turbulence, while in the phase turbulence regime,
where no phase slips occur,
 it has usually been considered that
 only the phase $\Phi(x,t)$ is active and the amplitude is effectively slaved
 to it \cite{Aranson_Kramer-RMP2002}. Below we show, among other things,
that this difference between  amplitude turbulence and phase turbulence
is reflected in the structure of the physical Lyapunov modes, thus rooting it
on a firm basis.

\subsection{Amplitude turbulence}  \label{sec:CGLAT}

We first focus on the amplitude turbulence regime,
 in which the amplitude and the phase evolve chaotically
 on rather short time and length scales.
Specifically, we use $\alpha = 2.5$, $\beta = -2.0$,
 $L = 64$, and $k_{\rm cut} = 31 \cdot 2\pi/L$.
Data are recorded over a period of about $10^7$
 after a transient period of $2.5 \times 10^4$.

\begin{figure}[t]
  \includegraphics[width=\hsize,clip]{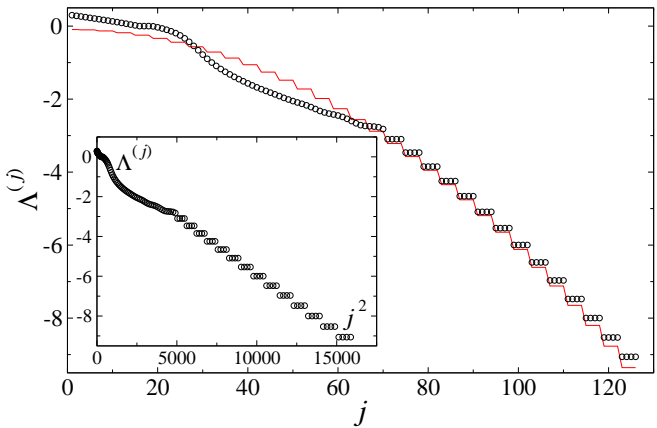}
  \caption{(color online). Lyapunov spectrum for the CGL amplitude turbulence. The stepwise structure starts at $j=71$. The red dashed line indicates $\Lambda^{(j)} = 1-k^2-2\expct{|W|^2}$ derived by Garnier and W\'ojcik for pure Fourier perturbations \cite{Garnier_Wojcik-PRL2006}. Inset: the same data against $j^2$. The Kaplan-Yorke dimension in this case is about $28.6$. }
  \label{fig:CGLATLyapSpec}
\end{figure}%

\begin{figure}[t]
  \includegraphics[width=\hsize,clip]{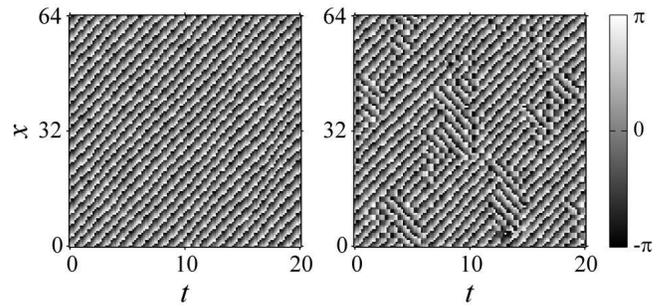}
  \caption{Spatiotemporal evolution of a typical CLV in the spurious region ($j = 78$) for the CGL amplitude turbulence. The phase component $\arg \delta W^{(j)}$ is plotted. The two plots are from the same trajectory but at two distant periods of time, showing pure upward traveling waves and patches of upward and downward waves, respectively.}
  \label{fig:CGLATVectorStructures}
\end{figure}

\begin{figure}[t]
  \includegraphics[width=\hsize,clip]{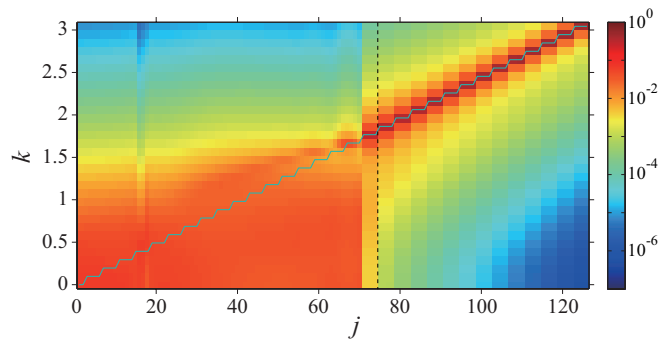}
  \caption{(color online). Spatial power spectrum $S_j(k)$ of the CLVs for the CGL amplitude turbulence, shown as a function of the wave number $k$ and the Lyapunov index $j$. Given that $S_j(k)$ and $S_j(-k)$ are statistically equivalent, their average is shown in the figure. The black dashed line separates the physical and spurious regions at $N_{\rm ph}=74$. The blue solid line indicates $k = [(j+1)/4] 2\pi/L$.}
  \label{fig:CGLATPowSpec}
\end{figure}

Figure \ref{fig:CGLATLyapSpec} shows the Lyapunov spectrum
 in the amplitude turbulence regime.
A stepwise region is present, as in the KS equation,
 but here the multiplicity of each step is four
 and the spatiotemporal evolution of CLVs in the stepwise region
 exhibits traveling waves (\figref{fig:CGLATVectorStructures})
 instead of the stationary wave found in the KS equation
 [\figref{fig:KSVectorStructures}(c)].
Traveling waves are the natural modes here, as can be readily seen
 from the linear stability analysis of the null solution $W(x,t)=0$,
 which gives $\exp[i(\pm kx-\omega t) + \Lambda t]$
 with $\Lambda = 1-k^2$ and $\omega = \alpha k^2$.
This makes the modes with the wave number $k$ and $-k$ distinguishable,
 albeit degenerate,
 and thus leads to the additional multiplicity of two
 in the CGL amplitude turbulence.
Indeed, each CLV in the stepwise region
 shows a sharp peak in the power spectrum,
 located at the trivially determined wave number of the $j$th traveling wave,
 namely at $k = [(j+1)/4] 2\pi/L$ (\figref{fig:CGLATPowSpec}).
The values of the Lyapunov exponents therefore decrease
 as $\Lambda^{(j)} \approx -k^2 \sim -j^2$ for large $j$
 (inset of \figref{fig:CGLATLyapSpec}).
A better estimate was obtained
 by Garnier and W\'ojcik \cite{Garnier_Wojcik-PRL2006},
 who derived $\Lambda^{(j)} = 1-k^2-2\expct{|W|^2}$
 for these stepwise exponents (red dashed line in \figref{fig:CGLATLyapSpec})
 assuming perturbations of pure Fourier modes.
In passing, the degeneracy of $k$ and $-k$ modes implies that
 these two modes can be in fact mixed up in a single vector;
 their CLVs are either in the pure $k$ mode,
 in the pure $-k$ mode, or patches of the two,
 changing their states very slowly in time
 (\figref{fig:CGLATVectorStructures}).

\begin{figure}[t]
  \includegraphics[width=\hsize,clip]{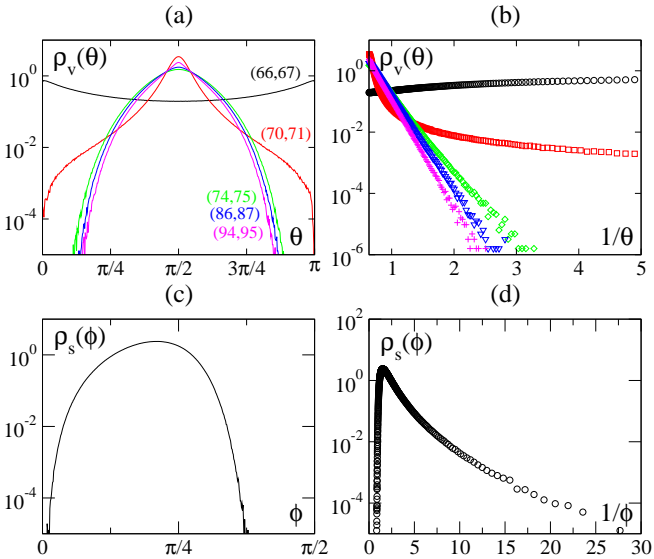}
  \caption{(color online). Distributions of the angle between neighboring CLVs (a,b) and between the physical and spurious subspaces (c,d) in the CGL amplitude turbulence. (a) Distribution function $\rho_{\rm v}(\theta)$ for CLV pairs $(66,67), (70,71), (74,75), (86,87), (94,95)$ from upper right to lower left. (b) $\rho_{\rm v}(\theta)$ versus $1/\theta$ for the data shown in (a). The ordinate is averaged over both sides of the distribution. (c) Angle distribution $\rho_{\rm s}(\phi)$ between the physical subspace spanned by CLVs of indices $1 \leq j \leq N_{\rm ph} = 74$ and the spurious one formed by the remaining CLVs. (d) The same data shown against $1/\phi$.}
  \label{fig:CGLATAngleDist}
\end{figure}%

In spite of these differences, the spurious modes
 in the amplitude turbulence are again hyperbolically isolated
 from any other modes.
Figure \ref{fig:CGLATAngleDist}(a) shows the angle distribution
 $\rho_{\rm v}(\theta)$ for pairs of the neighboring Lyapunov modes.
Similarly to the KS equation, the functional form of $\rho_{\rm v}(\theta)$
 sharply changes at the threshold located at the pair $(j,j+1)$
 with $j = N_{\rm ph} = 74$,
 above which the distributions are strictly bounded.
The tail of the distribution near the threshold exhibits
 the same critical decay $\rho_{\rm v}(\theta) \sim \exp(-\const/\theta)$
 within our numerical precision [\figref{fig:CGLATAngleDist}(b)].
The absence of tangencies is also checked
 from the distribution of the angle $\phi$
 between the physical and spurious subspaces
 spanned by the vectors $1 \leq j \leq N_{\rm ph}$ and $N_{\rm ph} < j$,
 respectively [\figref{fig:CGLATAngleDist}(c)].
The angle distribution $\rho_{\rm s}(\phi)$ is indeed bounded
 away from $\phi=0$.
Though the subspace angle $\phi$ is smaller than
 the angle $\theta$ for any pair of CLVs including a spurious mode,
 $\rho_{\rm s}(\phi)$ eventually decays as $\exp(-\const/\phi)$
 for small $\phi$ [\figref{fig:CGLATAngleDist}(d)],
 confirming the absence of any tangencies.
The spurious modes are therefore isolated from any other modes
 outside the step they belong to in the Lyapunov spectrum,
 and from any linear combination of the physical modes,
 in the same way as for the KS equation.

\begin{figure}[t]
 \includegraphics[width=\hsize,clip]{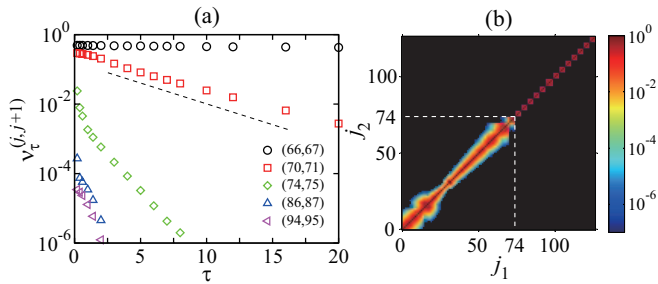}
 \caption{(color online). Violation of DOS in the CGL amplitude turbulence. (a) Time fraction $\nu_\tau^{(j,j+1)}$ of the DOS violation of the neighbor exponents as a function of $\tau$. The dashed line is a guide for the eyes showing an exponential decay. (b) $\nu^{(j_1,j_2)}_\tau$ for arbitrary pairs with $\tau = 16$. The black color indicates $\nu^{(j_1,j_2)}_\tau = 0$, i.e., hyperbolically isolated pairs. The white dashed lines indicate the threshold $N_{\rm ph} = 74$ between the physical and spurious modes.}
 \label{fig:CGLATDosViolation}
\end{figure}%

Concerning the DOS violation,
 though the time fraction $\nu_\tau^{(j_1,j_2)}$ measured with small $\tau$
 is not zero even in the spurious region,
 the same threshold $N_{\rm ph} = 74$ is found when
 we consider larger values of $\tau$ [\figref{fig:CGLATDosViolation}(a)].
Contrary to what happens for $j_1,j_2 \leq N_{\rm ph}$,
 $\nu^{(j_1,j_2)}_\tau$ for $j_2 > N_{\rm ph}$
 decreases faster than exponentially,
 though exponential decay may persist
 up to the largest $\tau$ explorable near the threshold.
This decay faster than exponential
 implies the existence of a finite $\tau$ beyond which
 $\nu^{(j_1,j_2)}_\tau$ is zero.
With such large $\tau$ values, every quartet of the spurious modes
 is indeed hyperbolically isolated from all the other modes,
 while the physical modes including those in the first apparent step
 are connected to the neighbors [\figref{fig:CGLATDosViolation}(b)].
All these results are consistent with what we have found
 for the KS equation and determine exactly the number of the physical modes,
 here at $N_{\rm ph} = 74$.

\subsection{Phase turbulence}

We now turn our attention to the phase-turbulence regime
 [\figref{fig:CGLSnapshots}(a)],
 in which the dynamics is believed
 to be governed by the phase variable \cite{Aranson_Kramer-RMP2002}.
We choose here $\alpha = 2.0$, $\beta = -0.91$, and $L = 96$.
The cutoff wave number is $k_{\rm cut} = 31 \cdot 2\pi/L$
 unless otherwise indicated.
In this phase-turbulence regime, the global phase difference is conserved
 because of the absence of phase slips.
Here we set it to be zero, in which case
phase dynamics is well-described by the KS equation, at least close to
the Benjamin-Feir line \cite{Aranson_Kramer-RMP2002}.
Data are recorded over a period of about $10^6$
 after a transient period of $5 \times 10^4$ is discarded.

\begin{figure}[t]
 \includegraphics[width=\hsize,clip]{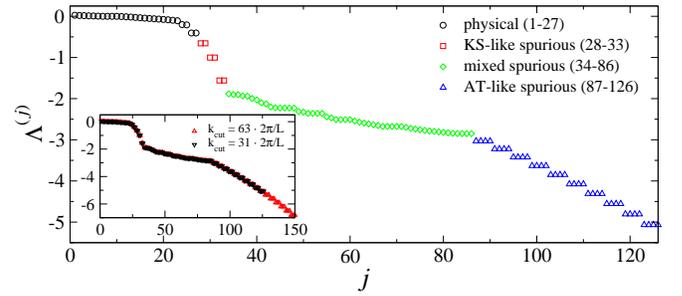}
  \caption{(color online). Lyapunov spectrum $\Lambda^{(j)}$ for the CGL phase turbulence. Different symbols correspond to different regions of the Lyapunov modes (see text). The Kaplan-Yorke dimension in this case is about $14.6$. Inset: Lyapunov spectra with different spatial resolutions.}
 \label{fig:CGLPTLyapSpec}
\end{figure}%

\begin{figure}[t]
 \includegraphics[width=\hsize,clip]{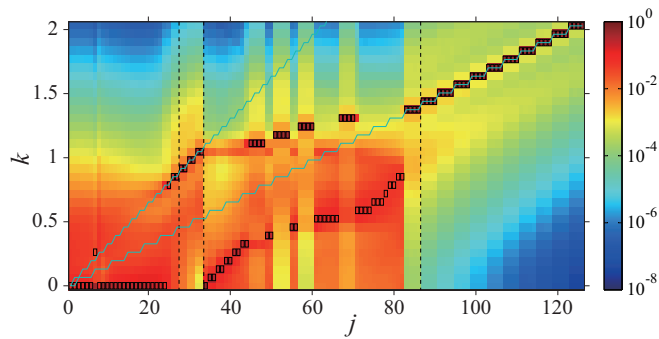}
  \caption{(color online). Spatial power spectrum $S_j(k)$ of the CLVs for the CGL phase turbulence, shown as a function of the wave number $k$ and the Lyapunov index $j$. Given that $S_j(k)$ and $S_j(-k)$ are statistically equivalent, their average is shown in the figure. The peak wave number for each $j$ is indicated by the black rectangles. The black dashed lines separate the four regions of the Lyapunov modes. The blue solid lines indicate $k = [j/2] 2\pi/L$ (upper line) and $k = [(j+1)/4] 2\pi/L$ (lower line), which are the trivial peak wave numbers found for the spurious modes in the KS equation and in the CGL amplitude turbulence, respectively.}
 \label{fig:CGLPTPowSpec}
\end{figure}%

In this regime, the Lyapunov spectrum
 shown in \figref{fig:CGLPTLyapSpec} reveals an interesting structure;
 now there are much less physical modes (black circles)
 than in the amplitude turbulence and
 they are followed by three distinct regions of spurious modes.
The modes in the first spurious region resemble those in the KS equation
 (red squares), while in the last part of the spectrum
 we find a structure similar to that in the CGL amplitude turbulence
 (blue triangles).
In between, the spurious modes appear to be mixed up
 with the modes that would take part in the physical region
 in the amplitude turbulence regime,
 as suggested from the rather smooth structure
 of the spectrum (green diamonds).
We shall therefore refer to these spurious modes as KS-like,
 amplitude-turbulence-like (AT-like), and mixed, respectively,
 as will be justified later.
The exact indices for all the thresholds will be given as well
 from their hyperbolicity properties.
Increasing the spatial resolution merely results in adding
 further AT-like spurious modes at the end of the spectrum
 without changing the essential features
 of the existing modes (inset of \figref{fig:CGLPTLyapSpec}).

Figure \ref{fig:CGLPTPowSpec} shows the spatial power spectra
 of the associated CLVs.
This justifies the above classification of the spurious modes;
 in the physical region the power spectrum of the CLVs
 reflects that of the trajectory (not shown) and thus
 it hardly changes with varying index,
 except for those associated with null Lyapunov exponents
 or too close to the threshold.
For the KS-like and AT-like spurious modes,
 the CLVs are nearly sinusoidal,
 as indicated by a sharp peak in their power spectrum.
Their peak wave number $k_{\rm peak}$ is determined
 by the same trivial geometrical rules as those for the KS equation
 and for the amplitude turbulence, respectively:
 $k_{\rm peak} = [j/2] 2\pi/L$ for the former
 and $k_{\rm peak} = [(j+1)/4] 2\pi/L$ for the latter
 (blue solid lines in \figref{fig:CGLPTPowSpec}).
In contrast, the mixed spurious region consists of two branches;
 one continues from the KS-like spurious modes,
 corresponding to the upper branch in the peak wave number,
 and the other continues from the physical modes, forming the lower branch.
The two branches finally merge,
 marking the start of the AT-like spurious region.

\begin{figure}[t]
 \includegraphics[width=\hsize,clip]{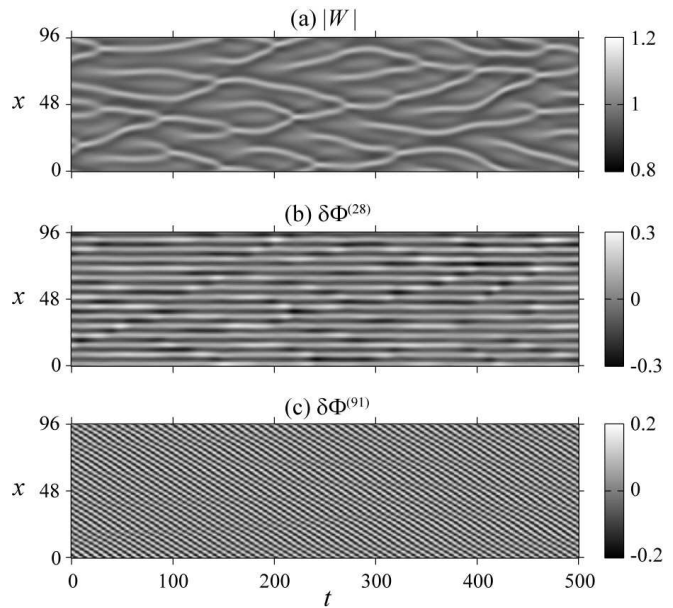}
 \caption{Spatiotemporal evolution of a typical KS-like and AT-like mode in the CGL phase turbulence. (a) Amplitude $|W|$ of the dynamics. (b,c) Phase component $\delta\Phi^{(j)}$ of the CLV for the KS-like spurious mode at $j=28$ (b) and for the AT-like one at $j=91$ (c).}
 \label{fig:CGLPTVectorStructures}
\end{figure}%

The correspondence of these modes to the spurious modes of the KS equation
 and the CGL amplitude turbulence
 is not found only in the peak wave number.
Figure \ref{fig:CGLPTVectorStructures} displays the spatiotemporal structure
 of a typical KS-like and AT-like spurious mode,
 showing specifically the local phase shift $\delta\Phi^{(j)}$
 onto the phase field $\Phi$ [\eqref{eq:DefinitionCGLPhase}]
 induced by the corresponding CLV.
It is defined by the relation
\begin{equation}
 \delta W^{(j)} = [\delta A^{(j)} + \ri (A\delta\Phi^{(j)} + \beta\delta A^{(j)})]\re^{\ri(\Phi+\beta\ln A)},  \label{eq:CGLVectorComponents}
\end{equation}
 obtained with $A(x,t) \equiv |W(x,t)|$ and
 $W = A\re^{\ri(\Phi+\beta\ln A)}$,
 and is a natural counterpart of the vector component $\delta u^{(j)}$
 for the KS equation.
In this phase shift field,
 the KS-like spurious modes [\figref{fig:CGLPTVectorStructures}(b)]
 reveal essentially stationary wave structure
 with amplitude modulated by the dynamics,
 in the same way as those for the KS equation
 [\figref{fig:KSVectorStructures}(c)].
Similarly, the same traveling wave structure
 as in the amplitude turbulence regime
 (bottom insets of \figref{fig:CGLATLyapSpec})
 is found for the AT-like spurious modes
 [\figref{fig:CGLPTVectorStructures}(c)].

\begin{figure}[t]
 \includegraphics[width=\hsize,clip]{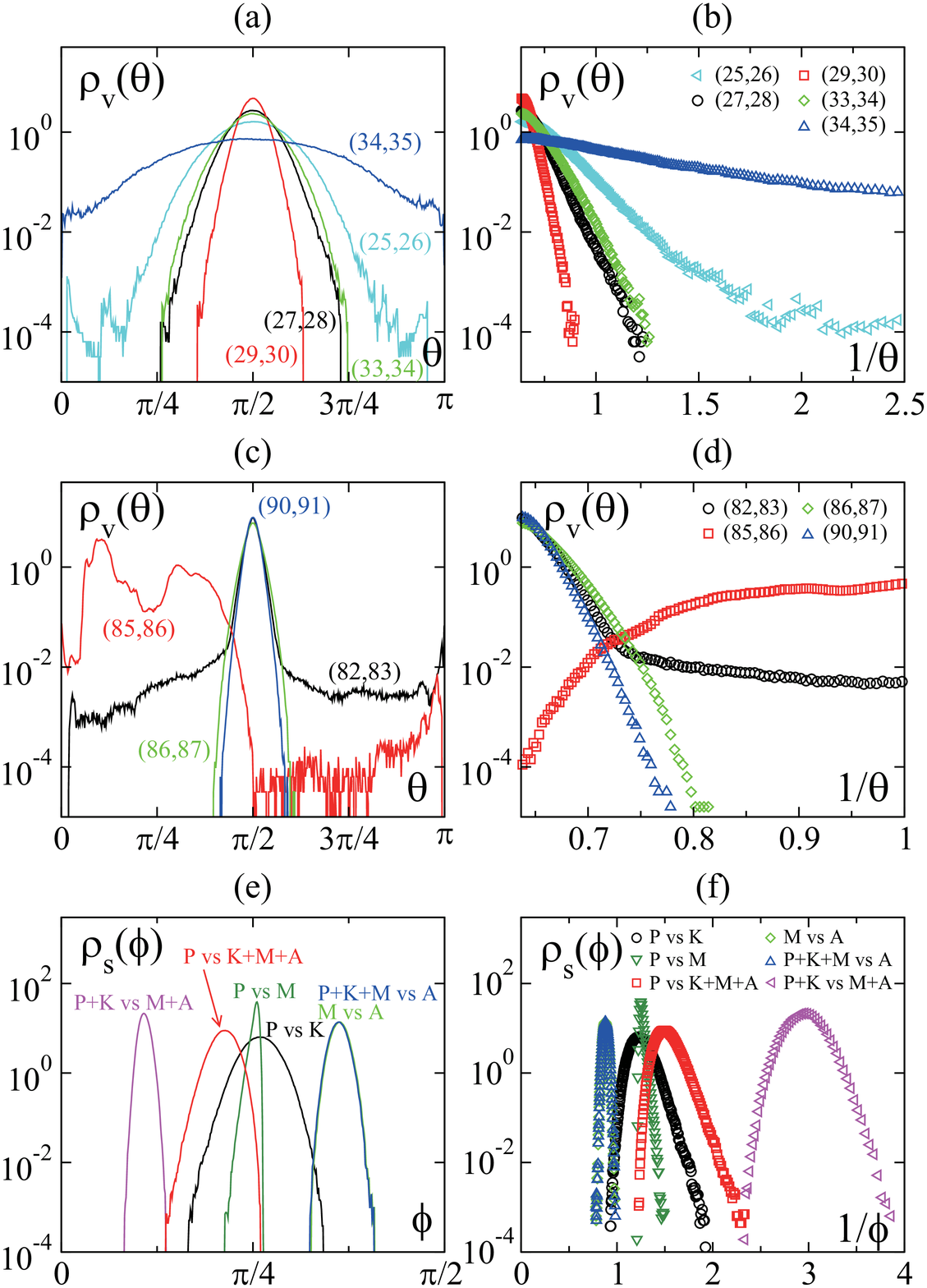}
 \caption{(color online). Distributions of the angle between neighboring CLVs (a-d) and between pairs of subspaces (e,f) in the CGL phase turbulence. (a) Distribution function $\rho_{\rm v}(\theta)$ for CLV pairs $(34,35), (25,26), (33,34), (27,28), (29,30)$ from outermost to innermost. They determine the thresholds for the physical, KS-like, and mixed regions. (b) $\rho_{\rm v}(\theta)$ versus $1/\theta$ for the data shown in (a). The ordinate is averaged over both sides of the distribution. (c) $\rho_{\rm v}(\theta)$ for CLV pairs $(82,83), (85,86), (86,87), (90,91)$ as indicated by the labels. They are relevant to the threshold between the mixed and AT-like regions. (d) $\rho_{\rm v}(\theta)$ versus $1/\theta$ for the data shown in (c). The ordinate is averaged over both sides of the distribution. (e) Subspace angle distributions $\rho_{\rm s}(\phi)$. The abbreviations P, K, M, and A refer to the physical modes ($1 \leq j \leq 27$), the KS-like spurious modes ($28 \leq j \leq 33$), the mixed spurious modes ($34 \leq j \leq 86$), and the AT-like spurious modes ($87 \leq j \leq 126$), respectively. The rightmost two curves overlap almost perfectly. (f) The same data as (e) shown against $1/\phi$.}
 \label{fig:CGLPTAngleDist}
\end{figure}%

In spite of this rather complicated internal structure of the spurious region,
 the essential feature of the spurious modes remains the same;
 they are all hyperbolically isolated from the physical modes.
Figure \ref{fig:CGLPTAngleDist}(a-d) shows the angle distributions
 for pairs of neighboring Lyapunov modes near the thresholds.
For the threshold between the physical and KS-like spurious modes,
 the first two visible doublets
 in the Lyapunov spectrum ($24 \leq j \leq 27$) belong in fact
 to the physical region
 [see the pair $(25,26)$ in \figref{fig:CGLPTAngleDist}(a,b)],
 whereas the first KS-like spurious mode is found at $j = 28$
 [pair $(27,28)$ in (a,b)].
The KS-like spurious modes are hyperbolic with respect to
 any other modes outside the step
 including other KS-like doublets [pair $(29,30)$ in (a,b)]
 and the mixed modes [pair $(33,34)$ in (a,b)].
In contrast, the mixed spurious modes,
 despite the absence of tangencies with any physical, KS-like,
 and AT-like modes, do have tangencies inside the mixed region
 [pair $(34,35)$ in (a,b) and pairs $(82,83), (85,86)$ in (c,d)].
The asymmetry (with respect to $\pi/2$) of some of the distributions
 is due to the finite length of the simulations,
 indicating that correlations relax very slowly
 for these pairs of CLVs.
In the AT-like spurious region,
 every quartet is again isolated from any other modes
 including other quartets in the AT-like region
 [pairs $(86,87), (90,91)$ in (c,d)].
From all these observations,
 we precisely determine the number of the physical modes
 at $N_{\rm ph} = 27$,
 as well as the index of all the internal thresholds of the spurious region
 as illustrated in \figref{fig:CGLPTLyapSpec}.

\begin{figure}[t]
 \includegraphics[width=\hsize,clip]{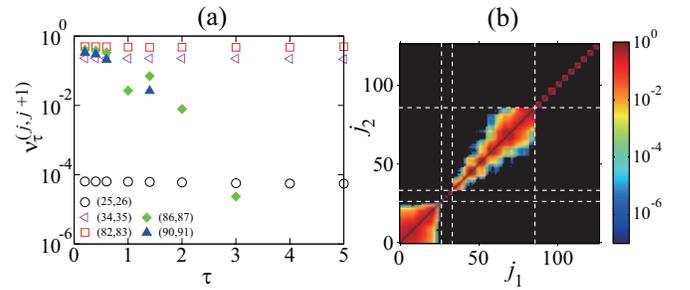}
 \caption{(color online). Violation of DOS in the CGL phase turbulence. (a) Time fraction $\nu_\tau^{(j,j+1)}$ of the DOS violation for pairs of neighboring Lyapunov exponents as a function of $\tau$. Pairs with and without tangencies in the angle distributions [\figref{fig:CGLPTAngleDist}(a,b)] are denoted by open and solid symbols, respectively. All pairs in the KS-like spurious region ($28 \leq j \leq 33$ excluding those in the same doublet) perfectly satisfy DOS, i.e., $\nu^{(j_1,j_2)}_\tau = 0$, even with such a small value of $\tau$ as $0.2$. (b) $\nu^{(j_1,j_2)}_\tau$ for arbitrary pairs with $\tau = 4$. The black color indicates $\nu^{(j_1,j_2)}_\tau = 0$, i.e., hyperbolically isolated pairs. The white dashed lines indicate the thresholds of the four regions.}
 \label{fig:CGLPTDosViolation}
\end{figure}%

The violation of DOS corroborates this view (\figref{fig:CGLPTDosViolation}).
Though the time fraction of the DOS violation $\nu_\tau^{(j_1,j_2)}$ is
 found to decrease sometimes not monotonically with respect to $\tau$
 for mixed spurious modes,
 presumably related to their long-lasting correlation,
 $\nu_\tau^{(j_1,j_2)}$ decays faster than exponentially for any pair
 which lacks tangencies in the angle distributions
 [\figref{fig:CGLPTDosViolation}(a)].
The matrix representation of $\nu_\tau^{(j_1,j_2)}$
 with a sufficiently large $\tau$, which is only $\tau = 4$ here,
 clearly shows that all KS-like doublets and AT-like quartets
 are completely isolated, while the mixed spurious modes are
 densely connected inside the mixed region but with none of the modes outside
 [\figref{fig:CGLPTDosViolation}(b)].

Similarly to the cases studied above, 
 not only pairs of single Lyapunov modes, but any linear combination of them
 is strictly hyperbolic if they are chosen from different regions.
This is best demonstrated by the angle distributions
 for subspaces spanning the four regions
 [\figref{fig:CGLPTAngleDist}(e,f)].
Subspaces formed by the Lyapunov modes spanning different regions
 have always nonzero angles between each other.
The angle distribution $\rho_{\rm s}(\phi)$ shows the critical decay
 with essential singularity, i.e.,
 $\rho_{\rm s}(\phi) \sim \exp(-\const/\phi)$,
 or decays even faster [\figref{fig:CGLPTAngleDist}(f)].
In particular, all the spurious modes, including mixed ones,
 do not have any tangency with any of the physical modes
 and their Lyapunov exponents are always negative.
The two essential properties of the spurious modes are thus
 satisfied by all of them.
This implies that
 they exponentially decay to zero without interacting with any physical modes,
 though some spurious modes are internally entangled,
 and thus they are not expected to affect the dynamics in the global attractor.

\begin{figure}[t]
 \includegraphics[width=\hsize,clip]{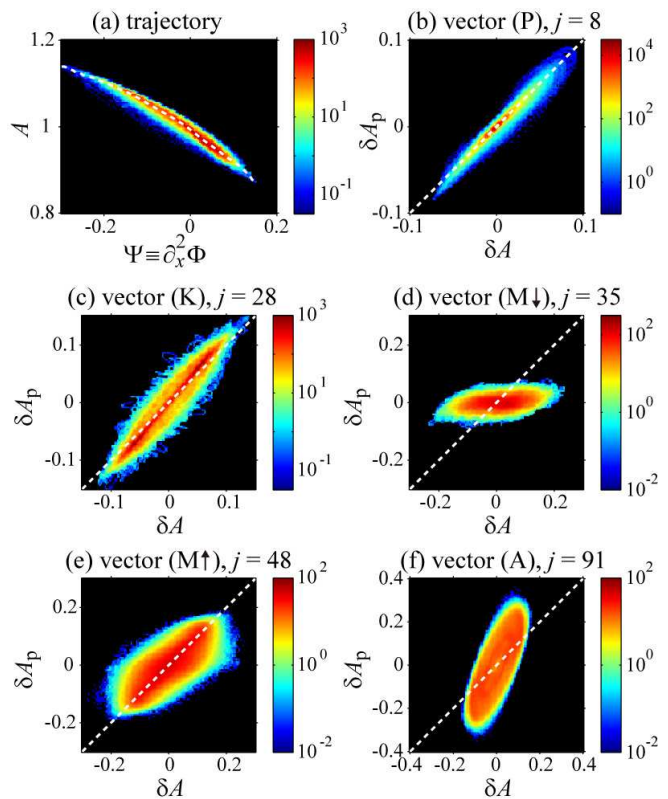}
 \caption{(color online). Density plots showing the structure of the trajectory (a) and of the CLVs (b-f) in the CGL phase turbulence. (a) Density plot for the amplitude $A$ and the phase curvature $\Psi$ of the trajectory. The white dashed curve indicates the averaged value of $A$ for each bin of $\Psi$. (b-f) Density plots for the CLVs, comparing the actual amplitude component $\delta A$ with that expected from the assumption of the slaved amplitude, $\delta A_{\rm p}$ [\eqref{eq:SlavedAmplitudeCLV}]. The white dashed lines indicate $\delta A_{\rm p} = \delta A$. The abbreviations P, K, M, and A denote a physical, KS-like, mixed, and AT-like mode, respectively. The upward and downward arrows next to M denote the upper and lower branches, respectively, in the plot of the peak wave number [\figref{fig:CGLPTLyapSpec}(b)]. All the data shown here are recorded during two periods of length $10^3$ separated by an interval of roughly $10^6$.}
 \label{fig:CGLPTVectorStructures2}
\end{figure}

These results for the amplitude turbulence and the phase turbulence
 lead us to interpret that amplitude degrees of freedom
 are slaved in the phase turbulence regime,
 and this produces the mixed spurious region in place,
 being mixed with the KS-like spurious modes.
To test this speculation, we compare the structure of the CLVs
 with that expected under the assumption of the amplitude slaved by the phase.
The dynamics in the phase turbulence regime indicates that
 the amplitude $A(x,t) \equiv |W(x,t)|$ is almost uniquely determined
 by the local phase curvature $\Psi(x,t) \equiv \p^2 \Phi/\p x^2$, i.e.,
\begin{equation}
 A(x,t) \approx A(\Psi(x,t)),  \label{eq:SlavedAmplitude}
\end{equation}
 as shown in \figref{fig:CGLPTVectorStructures2}(a).
This then implies the following amplitude-phase relation for the CLVs,
 as long as they result from the slaved amplitude:
\begin{equation}
 \delta A_{\rm p}^{(j)}(x,t) \approx \prt{A}{\Psi} \delta\Psi^{(j)}(x,t),  \label{eq:SlavedAmplitudeCLV}
\end{equation}
 where $\delta A$ and $\delta \Psi$ are the vector components
 corresponding to $A$ and $\Psi$, obtained through
 \eqref{eq:CGLVectorComponents}
 and $\delta\Psi^{(j)} = \p^2\delta\Phi^{(j)}/\p x^2$,
 and the subscript ``p'' denotes the estimate
 under the assumption \pref{eq:SlavedAmplitude}.
Comparing $\delta A_{\rm p}^{(j)}$
 with the actual amplitude component $\delta A^{(j)}$,
 one can examine to what extent the assumption of the slaved amplitude
 holds for each Lyapunov mode.

The result is shown in \figref{fig:CGLPTVectorStructures2}(b-f),
 which displays the density plots of $\delta A_{\rm p}^{(j)}$
 versus $\delta A^{(j)}$
 for a typical CLV in each region of the Lyapunov spectrum.
For the CLVs in the physical region [\figref{fig:CGLPTVectorStructures2}(b)]
 the prediction of \eqref{eq:SlavedAmplitudeCLV} holds
 as accurately as the slaved amplitude assumption \pref{eq:SlavedAmplitude}
 does for the dynamics [\figref{fig:CGLPTVectorStructures2}(a)].
It is also the case, albeit with somewhat larger deviations,
 for the KS-like spurious modes [\figref{fig:CGLPTVectorStructures2}(c)].
This is quite natural because the physical modes should reflect
 the dynamics, which is governed by the phase here,
 and the KS-like spurious modes are present
 in the KS equation, which phenomenologically describes such phase instability.
On the other hand, the mixed spurious modes,
 which comprise degrees of freedom
 not obeying the slaved amplitude assumption,
 exhibit very different structures.
For those which belong to the lower branch in the peak wave number spectrum
 [\figref{fig:CGLPTVectorStructures2}(d)],
 the amplitude $\delta A$ varies widely,
 while the phase component stays essentially quiescent,
 so does $\delta A_{\rm p}$.
This implies that these modes comprise mostly the amplitude degrees of freedom,
 which underpins the view that
 independent amplitude degrees of freedom are lost and become spurious
 in this phase turbulence regime.
In contrast, the mixed spurious modes in the upper branch
 are still influenced by \eqref{eq:SlavedAmplitudeCLV}
 [\figref{fig:CGLPTVectorStructures2}(e)].
This indicates that these modes essentially originate
 from the KS-like modes,
 but the slaved amplitude assumption holds only very loosely,
 probably because of the mixture with the spurious amplitude degrees of freedom
 in the lower branch.
Finally, for the AT-like spurious modes
 the slaved amplitude assumption does not work any more
 [\figref{fig:CGLPTVectorStructures2}(f)].
The two amplitudes rather rotate in the $\delta A$-$\delta A_{\rm p}$ plane
 irrespective of the dynamics, which is in line with their trivial
 traveling wave structure [\figref{fig:CGLPTVectorStructures}(c)].

In summary, we have found that in the phase turbulence regime
 the tangent-space dynamics consists of the physical modes
 and the three sets of spurious modes,
 all of which are hyperbolically decoupled from any physical mode.
These physical modes essentially stem from the phase degrees of freedom
 of the CGL equation, unlike in the amplitude turbulence
 where both phase and amplitude participate in the physical modes.
The loss of the amplitude degrees of freedom considerably reduces
 the number of the physical modes and produces instead
 the mixed spurious modes.
This view is supported by the fact that
 the mixed spurious region starts at the exponent value
 close to $-2$ [\figref{fig:CGLPTLyapSpec}(a)],
 which is the stability of the amplitude
 of a single Ginzburg-Landau oscillator.
Our results, therefore, corroborate
 the qualitative picture of the phase turbulence
 that the dynamics is governed by the phase,
 whereas, practically, the amplitude modes
 do not serve as independent degrees of freedom
 as they are essentially slaved by the phase.
It should be noted, however, that the physical and spurious modes
 do \textit{not} correspond exactly to
 the phase and amplitude degrees of freedom,
 respectively, as can be seen from the weak violation
 of \eqsref{eq:SlavedAmplitude} and \pref{eq:SlavedAmplitudeCLV}
 in the dynamics and the physical modes
 [\figref{fig:CGLPTVectorStructures2}(a,b)].
This implies that the phase reduction is not exact in the phase turbulence,
 in line with results from past studies \cite{Sakaguchi-PTP1990}.
To extract a proper set of global variables
 describing the dynamics is a difficult problem,
 which will be briefly discussed in Sec.\ \ref{sec:discussion}.

\section{1D Lattice Systems} \label{sec:Lattice1d}

So far we have studied two generic dissipative PDEs
 and found the separation
 of a finite number of physical modes from the remaining spurious modes.
We have argued that the physical modes correspond
 to the active degrees of freedom necessary to describe the dynamics
 and that they probably provide a tangent-space representation
 of the inertial manifold.
Although the concept of the inertial manifold is of great importance
 in such PDE systems, the existence of such a smooth invariant manifold
 of dimension smaller than the phase space
 is not restricted to PDEs.
We therefore study spatially discrete systems in this section,
 one defined with discrete time (i.e., map)
 and another with continuous time (i.e., ordinary differential equation).

\subsection{1D Lattice of Coupled Tent Maps} \label{sec:CML1d}

First we consider the following
 one-dimensional lattice of diffusively-coupled tent maps:
\begin{align}
 x_i^{t+1} &= f(x_i^t) + \frac{K}{2}[ f(x_{i+1}^t) - 2f(x_i^t) + f(x_{i-1}^t)]  \notag\\
 &= \[ 1 + \frac{K}{2}\mathcal{D} \] f(x_i^t)  \label{eq:TentCML1}
\end{align}
 with $i = 1, 2, \cdots, L$, $f(x) = 1 - \mu |x|$,
 and the discrete Laplacian operator
\begin{equation}
 \mathcal{D}f(x_i^t) \equiv f(x_{i+1}^t) - 2f(x_i^t) + f(x_{i-1}^t).  \label{eq:DefinitionDiscreteLaplacian}
\end{equation}
In such systems, obviously, varying the spatial resolution does not make sense.
Instead, we increase the coupling constant $K$
 so that local dynamical variables $x_i^t$
 possess stronger correlations with their neighbors, possibly reducing
the number of effective degrees of freedom.

Here we use $\mu = 1.1$, $L = 256$, and PBC.
With this value of $\mu$, the local map exhibits four-band chaos.
We start from random initial conditions distributed
 only within one of the bands,
 and hence the coupled-map lattice shows period-4 collective behavior.
Data are recorded over roughly $5 \times 10^5$ time steps
 after a transient of nearly the same length.



\begin{figure}[t]
 \includegraphics[width=\hsize,clip]{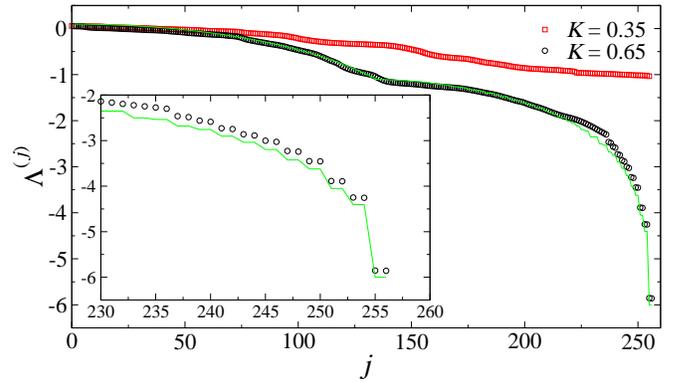}
 \caption{(color online). Lyapunov spectrum for the coupled tent maps \pref{eq:TentCML1} with varying coupling constant $K$. Inset: close-up of the end of the spectrum for $K=0.65$. The green lines show the spectrum estimated under the Fourier mode approximation (see text). The Kaplan-Yorke dimension for $K=0.65$ is about $39.9$.}
 \label{fig:TentCMLLyapSpec}
\end{figure}%
\begin{figure}[t]
 \includegraphics[width=\hsize,clip]{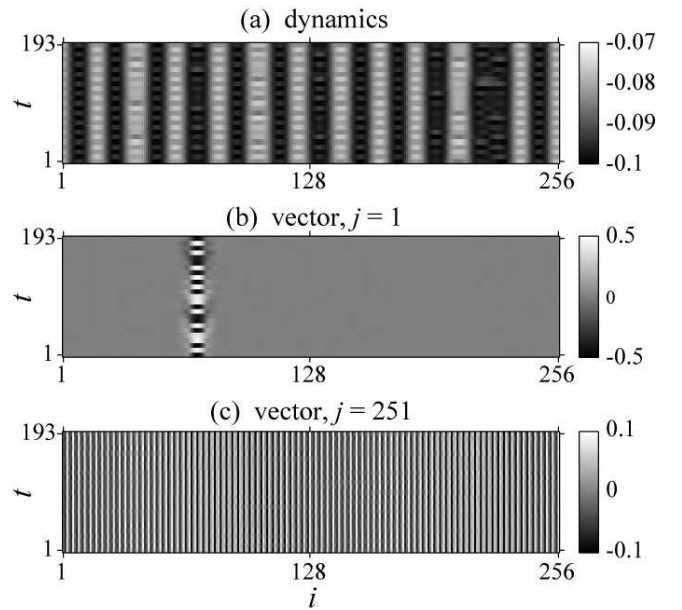}
 \caption{Spatiotemporal evolution of the dynamics $x_i^t$ (a), of a physical Lyapunov mode at $j=1$ (b), and of a spurious Lyapunov mode at $j=251$ (c) in the coupled tent maps with $K=0.65$. The spatial profiles are plotted every 8 time steps in order to capture the internal structure of the period-4 collective behavior.}
 \label{fig:TentCMLSnapshots}
\end{figure}%

Figure \ref{fig:TentCMLLyapSpec} shows the Lyapunov spectra
 of this coupled-map lattice for two values of the coupling constant $K$.
As expected, 
 while for moderate strengths of the coupling
 the Lyapunov spectrum does not show any splitting of the structure
 (e.g., red squares in \figref{fig:TentCMLLyapSpec}),
 strengthening the coupling induces a stepwise structure
 at the end of the Lyapunov spectrum
 [$K=0.65$ in \figref{fig:TentCMLLyapSpec} (black circles)],
 similarly to what happens in PDEs.

Concerning the structure of the CLVs,
 the Lyapunov modes in the smooth region,
 particularly at the beginning of the spectrum,
 tend to be localized at one of the antinodes (or a few of them)
 of the wavy structure of the dynamics
 [\figref{fig:TentCMLSnapshots}(a,b)].
Over long time scales the weight shifts from an antinode to another
 (or from some to others).
In constrast, those in the stepwise region
 exhibit stationary wave structures
 [\figref{fig:TentCMLSnapshots}(c)]
 similar to the spurious modes in the KS equation
 [\figref{fig:KSVectorStructures}(c)].

\begin{figure}[t]
 \includegraphics[width=\hsize,clip]{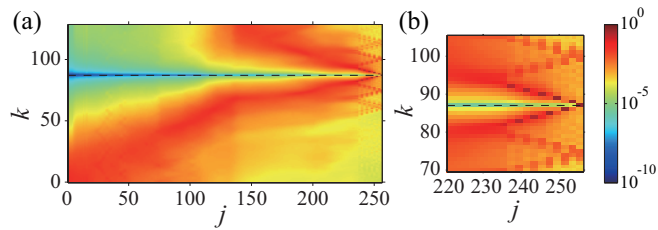}
 \caption{Spatial power spectrum of the CLVs for the tent coupled-map lattice with $K=0.65$, shown as a function of the (integer) wave number $k$ and the Lyapunov index $j$. The dashed line indicates $k=87$, which gives the smallest eigenvalue for the couling operator $1+(K/2)\mathcal{D}$ (see text). The panel (b) is a close-up of (a) near the peak wave numbers of the CLVs at the end of the Lyapunov spectrum.}
 \label{fig:TentCMLPowSpec}
\end{figure}%

A difference from the PDE systems
 is found however in the way the peak wave number of the spurious modes
 is determined.
Figure \ref{fig:TentCMLPowSpec} shows the spatial power spectra of the CLVs.
While in PDEs the peak wave number in the stepwise region
 increases monotonously according to trivial geometrical rules
 (Figs.\ \ref{fig:KSPowerSpectra} and \ref{fig:CGLPTPowSpec},
 except for the special case of the mixed spurious modes
 in the CGL phase turbulence),
 in the coupled-map lattice it forms two branches,
 one increasing and the other decreasing,
 which eventually join at $j=L$ at a certain wave number
 [dark red spots in \figref{fig:TentCMLPowSpec}(b)].
In fact, this is again a simple consequence
 of the nearly sinusoidal structure of the CLVs in the stepwise region.
The Fourier modes $\delta W_i = \sin(2\pi ik/L)$ and $\cos(2\pi ik/L)$
 with integer $k$
 are eigenvectors of the coupling operator $1+(K/2)\mathcal{D}$
 in \eqref{eq:TentCML1}.
Their associated eigenvalue is $1-K[1-\cos (2\pi k/L)]$,
 whose contribution to the Lyapunov exponent is
 $\log |1-K[1-\cos (2\pi k/L)]|$.
Since it is not a monotonous function of $k$,
 nor is the peak wave number in the power spectra.
For $K=0.65$, $k=87$ gives the eigenvalue closest to zero,
 and therefore is the peak wave number for the CLV
 of the smallest Lyapunov exponent
 (dashed line in \figref{fig:TentCMLPowSpec}).
Moreover, approximating the contribution from the local map $f(x)$
 to the Lyapunov exponent by $\log\mu$,
 we obtain $\Lambda = \log |1-K[1-\cos (2\pi k/L)]| + \log\mu$,
 which indeed captures the qualitative structure
 of the observed Lyapunov spectrum (\figref{fig:TentCMLLyapSpec}).

\begin{figure}[t]
 \includegraphics[width=\hsize,clip]{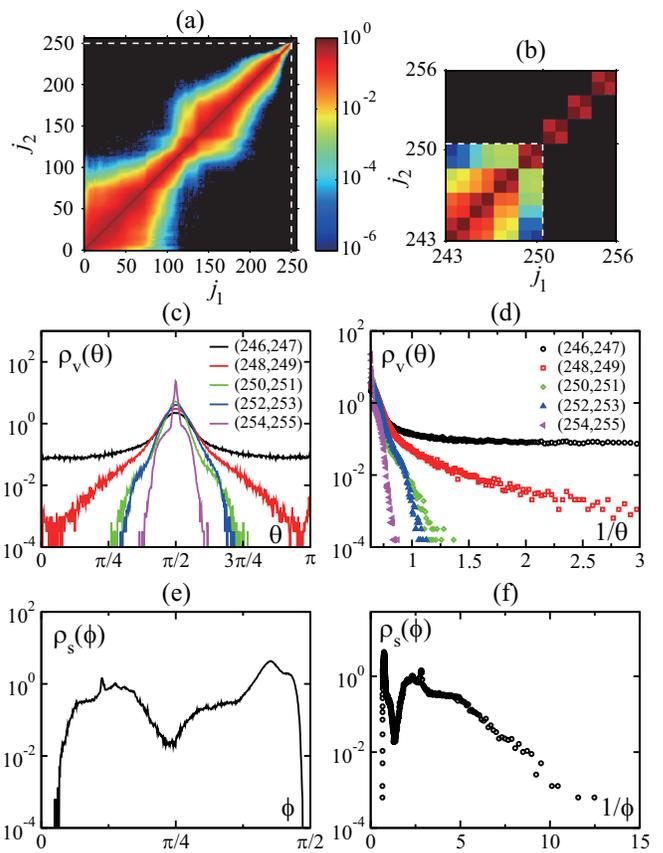}
  \caption{(color online). Hyperbolicity properties for the tent coupled-map lattice with $K=0.65$. (a) Time fraction $\nu^{(j_1,j_2)}_\tau$ of the DOS violation for arbitrary pairs with $\tau = 2$. The black color indicates $\nu^{(j_1,j_2)}_\tau = 0$, i.e., hyperbolically isolated pairs. The white dashed lines indicate the threshold $N_{\rm ph} = 250$ between the physical and spurious modes. (b) Close-up of (a) near the end of the spectrum. (c) Angle distributions $\rho_{\rm v}(\theta)$ for neighboring CLVs of indices $(246, 247), (248, 249), \cdots, (254, 255)$ from upper left to lower right. (d) $\rho_{\rm v}(\theta)$ against $1/\theta$. The ordinate is averaged over both sides of the distribution. (e) Angle distribution $\rho_{\rm s}(\phi)$ between the physical subspace ($1 \leq j \leq 250$) and the spurious one ($251 \leq j \leq 256$). (f) $\rho_{\rm s}(\phi)$ against $1/\phi$.}
 \label{fig:TentCMLHyperbolicity}
\end{figure}%

Despite this apparent difference from PDEs,
 we identify the same tangent-space decoupling
 into the physical and spurious modes in this coupled-map lattice.
Measuring the time fraction $\nu_\tau^{(j_1,j_2)}$ of the DOS violation,
 we find that the last three doublets of Lyapunov modes
 are hyperbolically isolated from all the other modes
 [\figref{fig:TentCMLHyperbolicity}(a,b)].
The splitting into the physical and spurious modes can also be confirmed
 from the angle distributions between CLVs
 [\figref{fig:TentCMLHyperbolicity}(c,d)],
 as well as that between the physical and spurious subspaces
 defined by the corresponding CLVs
 [\figref{fig:TentCMLHyperbolicity}(e,f)].
If the manifold of the physical modes is a local approximation
 of the inertial manifold as we speculate,
 our results imply that dissipative coupled map lattices may also have
 an inertial manifold, whose dimension is lower than that of the phase space.

\subsection{1D Lattice of Ginzburg-Landau Oscillators} \label{sec:LocGL1d}

Now we study another example of the spatially discrete dynamical systems,
 here defined with continous time.
Specifically, we take the system studied recently
 by Kuptsov and Parlitz \cite{Kuptsov_Parlitz-PRE2010},
 namely a one-dimensional lattice of CGL oscillators,
\begin{equation}
 \prt{W_i}{t} = W_i - (1+\ri\beta)|W_i|^2 W_i + (1+\ri\alpha)\frac{\mathcal{D}W_i}{h^2},
 \label{eq:DefinitionLocGL}
\end{equation}
 with complex variables $W_i(t)$ and $i = 1, 2, \cdots, L$.
This can be regarded as a spatial discretization of the CGL equation
 \pref{eq:DefinitionCGL},
 where the parameter $h$ denotes the lattice constant,
 and thus behaves similarly unless $h$ is too large.
For small $h$, the system is a good approximation of the CGL equation,
 and hence the decoupling of the physical and spurious modes takes place
 in the same manner as we have studied in Sec.\ \ref{sec:CGL1d}.
Increasing $h$,
 besides degrading the correspondence to the CGL equation,
 amounts to increasing the chain length $hL$
 and thus leads to an increase in the number of physical modes.
Moreover, Kuptsov and Parlitz \cite{Kuptsov_Parlitz-PRE2010}
 reported that, as the number of physical modes increases with $h$
 at the positive end of the Lyapunov spectrum,
 another set of modes appears at the negative end,
 which are hyperbolically connected within themselves
 and whose number also increases with $h$.
Isolated steps of the spurious modes being replaced from both ends,
 the authors argued that the two groups finally meet each other
 and the strict hyperbolic decoupling between them is lost for large $h$.
However, their claim is based solely on the observation
 of the structure of the CLVs
 and the time fraction of the DOS violation with a fixed $\tau$
 \cite{NoteKP,Kuptsov_Parlitz-PRE2010},
 which cannot fully determine hyperbolic decoupling.
Therefore, here we revisit this problem,
 performing a complete analysis
 to clarify the nature of the hyperbolic decoupling in this system,
 and explains this rather peculiar behavior reported by Kuptsov and Parlitz.

In the following, we take the same parameter values
 as Kuptsov and Parlitz \cite{Kuptsov_Parlitz-PRE2010},
 namely $\alpha = -2.0$ and $\beta = 3.0$
 which correspond to the amplitude turbulence regime
 (cf.\ Sec.\ \ref{sec:CGLAT}), and vary the value of $h$.
We fix the system size at $L=32$ and adopt PBC, $W_{i+L} = W_i$.
Kuptsov and Parlitz used instead a no-flux boundary condition
 in Ref.\ \cite{Kuptsov_Parlitz-PRE2010},
 but this difference in the boundary conditions intervenes
 practically only in the multiplicity of the stepwise structure
 of the Lyapunov spectrum, as we have seen in Sec.\ \ref{sec:KS1d}.
Time integration is performed
 with the fourth order Runge-Kutta method,
 over a period of about $2 \times 10^5$
 after a transient of $2.5 \times 10^4$ is discarded.

\begin{figure}[t]
 \includegraphics[width=\hsize,clip]{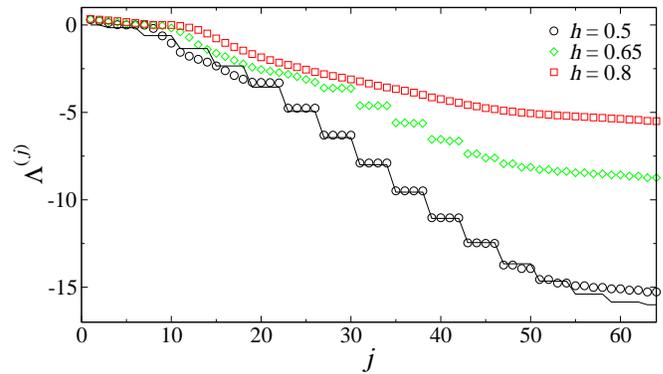}
 \caption{(color online). Lyapunov spectrum for the CGL lattice \pref{eq:DefinitionLocGL} with varying $h$. The numerically obtained spectra are shown by the symbols, while the black line indicates the spectrum of the contribution from the coupling term under the Fourier mode approximation for $h=0.5$ (see text). The Kaplan-Yorke dimension is about $8.8$, $11.6$, and $14.7$ for $h=0.5$, $0.65$, and $h=0.8$, respectively.}
 \label{fig:LocGLLyapSpec}
\end{figure}%

\begin{figure}[t]
 \includegraphics[width=\hsize,clip]{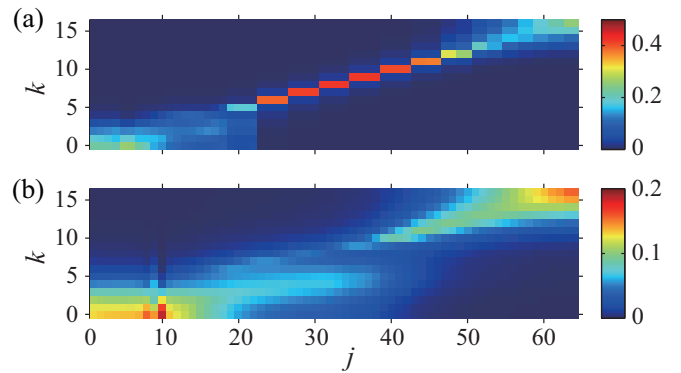}
 \caption{(color online). Spatial power spectrum $S_j(k)$ of the CLVs for the CGL lattice with $h=0.5$ (a) and $h=0.8$ (b), shown as a function of the (integer) wave number $k$ and the Lyapunov index $j$. Given that $S_j(k)$ and $S_j(-k)$ are statistically equivalent, their average is shown in the figure. Note that the color plots are in different linear scales for the sake of clearity.}
 \label{fig:LocGLPowSpec}
\end{figure}

Figure \ref{fig:LocGLLyapSpec} shows
 the Lyapunov spectra of the CGL lattice for varying $h$.
As reported in Ref.\ \cite{Kuptsov_Parlitz-PRE2010},
 for small $h$ the spectrum exhibits the stepwise structure
 (black circles),
 but it arises in the middle of the spectrum,
 as opposed to the CGL equation
 for which the steps continue up to the end of the spectrum.
In the CGL lattice, they are instead sandwiched by two smooth regions.
With increasing $h$, both of these smooth regions grow
 (green diamonds in \figref{fig:LocGLLyapSpec})
 and finally merge (red squares),
 replacing all the steps in the spectrum.
The power spectra of the CLVs in \figref{fig:LocGLPowSpec} show
 that the steps here are again due to the sinusoidal structure of the CLVs,
 as evidenced by the sharp peaks
 for the stepwise modes in \figref{fig:LocGLPowSpec}(a),
 while CLVs in the two smooth regions do not have obvious wavy structure.

\begin{figure}[t]
 \includegraphics[width=\hsize,clip]{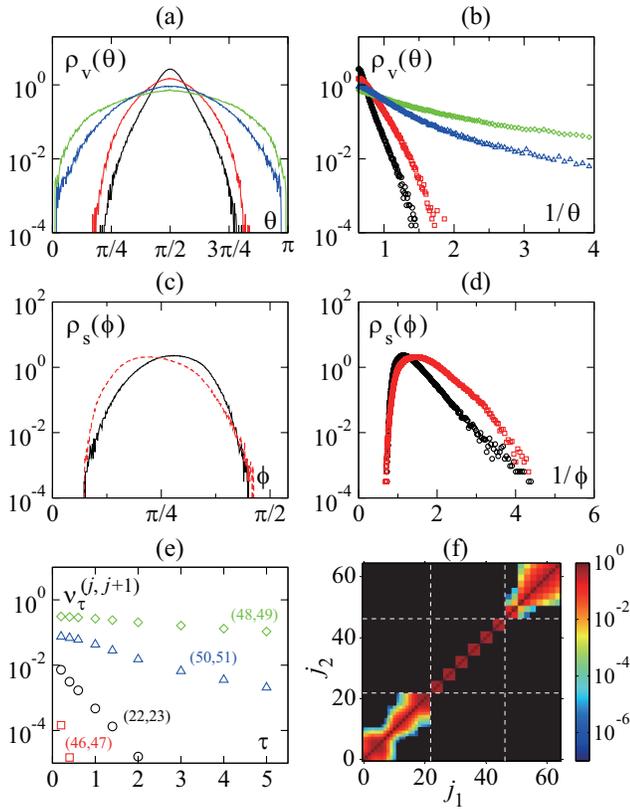}
 \caption{(color online). Hyperbolicity properties for the CGL lattice with $h=0.5$. (a) Angle distributions $\rho_{\rm v}(\theta)$ for neighboring CLVs of indices $(22, 23), (46, 47), (50, 51)$, and $(48,49)$ from innermost to outermost curve. (b) $\rho_{\rm v}(\theta)$ against $1/\theta$. The ordinate is averaged over both sides of the distribution. The order of the curves is the same as in (a) from lower left to upper right. (c) Angle distribution $\rho_{\rm s}(\phi)$ between the physical subspace ($1 \leq j \leq N_{\rm ph}$) and the spurious one ($N_{\rm ph}+1 \leq j$) with $N_{\rm ph}=22$ (black solid line). The red dashed line shows the angle distribution between the subspaces of indices $1 \leq j \leq 46$ and $47 \leq j \leq 64$, showing the hyperbolic isolation of the tangled spurious subspace. (d) The same plots as (c) but shown against $1/\phi$ (black circles for the physical and spurious subspaces, and red squares for the other). (e) Time fraction $\nu_\tau^{(j,j+1)}$ of the DOS violation for pairs of neighboring Lyapunov exponents as a function of $\tau$. (f) Time fraction $\nu^{(j_1,j_2)}_\tau$ of the DOS violation for arbitrary pairs with $\tau = 3$. The black color indicates $\nu^{(j_1,j_2)}_\tau = 0$, i.e., hyperbolically isolated pairs. The white dashed lines indicate the threshold $N_{\rm ph} = 22$ between the physical and spurious modes, and that between the physical and stepwise spurious modes $1 \leq j \leq 46$ and the tangled spurious modes $47 \leq j \leq 64$.}
 \label{fig:LocGLHyperbolicity1}
\end{figure}%

Analyzing hyperbolicity properties of this system
 reveals that, for small $h$, the Lyapunov spectrum splits into three regions,
 basically according to its stepwise structure
 (\figref{fig:LocGLHyperbolicity1}).
The first gap in the stepwise region already exhibits
 clear hyperbolic decoupling of the two modes at the edge,
 namely those of indices $22$ and $23$;
 for them the CLV angle distribution $\rho_{\rm v}(\theta)$ is bounded
 with a decay faster than $\exp(-\const/\theta)$ in the tail
 [black, innermost line and symbols in \figref{fig:LocGLHyperbolicity1}(a,b)],
 and the time fraction of the DOS violation $\nu_\tau^{(22,23)}$
 decreases faster than exponentially
 [black circles in \figref{fig:LocGLHyperbolicity1}(e)].
This fixes the number of physical modes at $N_{\rm ph} = 22$.
The decoupling holds similarly
 for any pair of CLVs sandwiching this threshold
 [\figref{fig:LocGLHyperbolicity1}(f)], as well as
 between the physical and spurious subspaces, defined at this index
 [black solid line and circles in \figref{fig:LocGLHyperbolicity1}(c,d)].

The spurious subspace in this system, however, consists of
 the central stepwise region and the smooth region
 at the end of the Lyapunov spectrum
 (\figref{fig:LocGLLyapSpec}).
In the former, quartets of Lyapunov modes are isolated
 from all the other modes [\figref{fig:LocGLHyperbolicity1}(f)]
 as is usually the case for the spurious modes,
 while in the latter, specifically here for indices larger than $46$,
 the spurious modes are internally connected
 [\figref{fig:LocGLHyperbolicity1}(a,b,e,f)]
 in a way similar to the mixed spurious modes in the CGL phase turbulence
 [see, e.g., \figref{fig:CGLPTDosViolation}(b)].
These are the modes found
 by Kuptsov and Parlitz \cite{Kuptsov_Parlitz-PRE2010},
 though in their work
 there remained ambiguity in the hyperbolicity properties
 as discussed above.
Here we confirm that these modes are indeed connected internally,
 but never with any physical or stepwise spurious mode,
 both from the DOS assay with varying $\tau$
 and from the CLV angle distribution
 [\figref{fig:LocGLHyperbolicity1}(a,b,e)].
For this property, we call these modes tangled spurious modes.
The subspace composed of them is also hyperbolically isolated
 from its complementary
 [red dashed line and squares in \figref{fig:CGLPTDosViolation}(c,d)].
One should be reminded that the difference between the tangled spurious modes
 and the physical ones lies in the sign of their Lyapunov exponents;
 while the exponents for the physical modes can take any sign,
 those for the spurious modes are always negative,
 and thus they decay without intervening in the dynamics
 even if they are internally connected.
The effective dimension of the system is therefore always
 given by the number of the physical modes, $N_{\rm ph}$,
 without contribution from the tangled spurious modes.

\begin{figure}[t]
 \includegraphics[width=\hsize,clip]{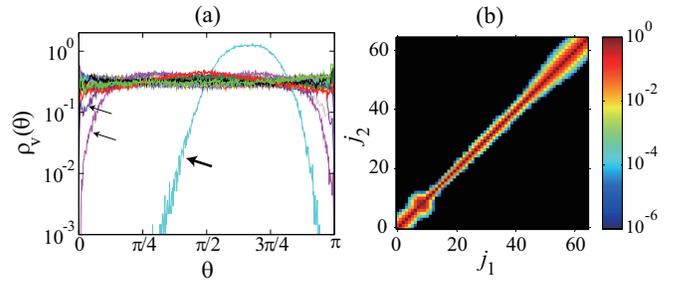}
 \caption{(color online). Hyperbolicity properties for the CGL lattice with $h=0.8$. (a) Angle distributions $\rho_{\rm v}(\theta)$ for all pairs of neighboring CLVs. Indicated by the arrows are the distributions for the pairs including either or both of the two Lyapunov modes with $\Lambda^{(j)} = 0$. All the distributions except the one for the two zero Lyapunov modes, shown by the bold arrow, have finite densities at $\theta = 0$ and $\pi$. (b) Time fraction $\nu^{(j_1,j_2)}_\tau$ of the DOS violation for arbitrary pairs with $\tau = 5$. The black color indicates $\nu^{(j_1,j_2)}_\tau = 0$, i.e., hyperbolically isolated pairs.}
 \label{fig:LocGLHyperbolicity2}
\end{figure}%

The appearance of the tangled spurious modes is in fact, again,
 due to the discrete Laplacian operator of the coupling term
 in \eqref{eq:DefinitionLocGL},
 but it takes place in a different manner as the tent coupled-map lattice
 because of the continuous time.
Here, if we take into account only the coupling term,
 its growth rate is simply given by the real part of its eigenvalues,
 specifically
 $\Lambda = -(2/h^2)[1-\cos(2\pi k/L)]$
 for the eigenmodes $\re^{2\pi\ri ki/L}$.
It is monotonic with respect to $|k|$, since $|k|$ varies only up to $L/2$,
 and hence so is the peak wave number of the CLV power spectra
 in the stepwise region of the Lyapunov spectrum
 (\figref{fig:LocGLPowSpec}, to be compared with \figref{fig:TentCMLPowSpec}).
The values of the Lyapunov exponent in this region are well approximated
 by $\Lambda$ given above with the corresponding $k$
 (black line in \figref{fig:LocGLLyapSpec}).
By contrast, for larger $k$ or $j$,
 the gap in $\Lambda$ between different $k$ shrinks,
 as opposed to the case of the continuous Laplacian operator.
This facilitates the order exchange of the Lyapunov exponents,
 i.e., the violation of the DOS, due to their intrinsic fluctuations
 from the dynamics, and thus produces the tangled spurious modes.

With increasing $h$, both physical and tangled spurious modes
 increase their numbers, replacing the steps of the spurious modes in between
 (cf., \figref{fig:LocGLLyapSpec}).
This finally gives rise to the coalescence of the two groups;
 as soon as the last physical mode and the first tangled spurious one
 encounter, all the modes are connected and thus become physical modes.
The case for $h=0.8$ is shown in \figref{fig:LocGLHyperbolicity2}.
All the CLV angle distributions have finite densities at $\theta = 0$ and $\pi$
 except for those concerning one or both of the two zero Lyapunov modes
 [\figref{fig:LocGLHyperbolicity2}(a)].
The overall connection of the Lyapunov modes is also confirmed
 form the violation of the DOS [\figref{fig:LocGLHyperbolicity2}(b)].
That is to say, with increasing $h$,
 the number of the physical modes,
 or the inertial manifold dimension as we conjecture, is suddenly doubled
 when the tangled spurious modes join the physical manifold.
Although such a large value of $h$ (around $0.7$ in this case)
 corresponds to a somewhat unphysical situation
 when we view the CGL lattice as an approximation to the original CGL equation,
 the physical implication and consequence of this ``crisis'' 
 remain to be clarified.

\section{2D Kuramoto-Sivashinsky equation}  \label{sec:KS2d}

\subsection{Definition and numerical scheme}

In the previous sections we have investigated the physical-spurious decoupling
 in 1D dissipative systems.
Given the generality of our finding, and also
 the general existence of a finite-dimensional inertial manifold
 in dissipative systems,
 one expects that the same decoupling should also take place in systems
 of higher dimensions.
This is indeed the case, as we shall demonstrate in this section
 for the 2D KS equation as a generic example.

The 2D KS equation is defined for a real-valued field $u(x,y,t)$
 as \cite{Note3}
\begin{equation}
 \prt{u}{t} = -\nabla^2 u - (\nabla^2)^2 u - \frac{1}{2}(\bnabla u)^2,
 \label{eq:DefinitionKS2d}
\end{equation}
 with $\bnabla \equiv (\p/\p x, \p/\p y)$,
 $x \in [0,L_x]$, and $y \in [0,L_y]$.
Here we adopt PBC $u(x,y,t) = u(x+L_x,y,t) = u(x,y+L_y,t)$
 with $L \equiv L_x = L_y = 24$ for the sake of simplicity.
For integration we use the same operator-splitting method
 as for the 1D KS equation and the pseudospectral method
 with $k_{\rm cut} = 10 \cdot 2\pi/L$,
 i.e., with $11 \times 11$ Fourier modes.
Integration is performed over a period of roughly $2 \times 10^5$
 after discarding a transient period of $1200$.

\subsection{Results}

\begin{figure}[t]
 \includegraphics[width=\hsize,clip]{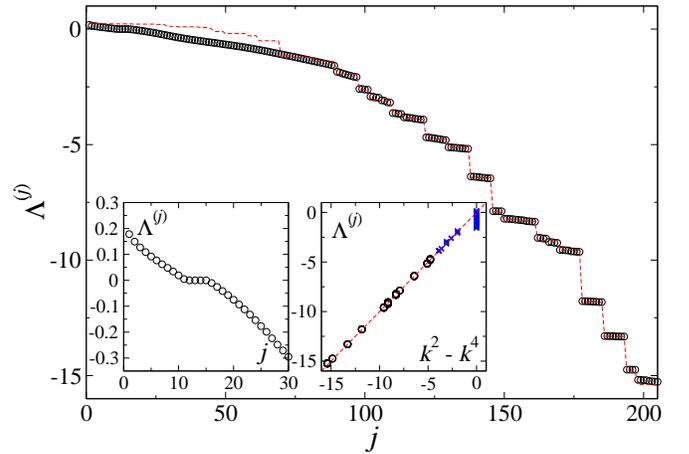}
  \caption{(color online). Lyapunov spectrum for the 2D KS equation. Red dashed line shows the values of $\Lambda^{(j)}$ estimated from the dominating wave number $(k_x,k_y)$ assigned trivially by the index (see text). Left inset: close-up of the beginning of the spectrum. Right inset: Lyapunov exponent $\Lambda^{(j)}$ vs $k^2-k^4$, where $k^2 \equiv k_x^2 + k_y^2$ is the square of the peak wave number for the power spectrum of the $j$th CLV (see \figref{fig:KS2dPowSpec}). Blue crosses and black circles correspond to the physical and spurious modes, respectively. Red dashed line indicates $\Lambda^{(j)} = k^2-k^4$. The Kaplan-Yorke dimension in this case is about $25.1$.}
 \label{fig:KS2dLyapSpec}
\end{figure}%

\begin{figure}[t]
 \includegraphics[width=\hsize,clip]{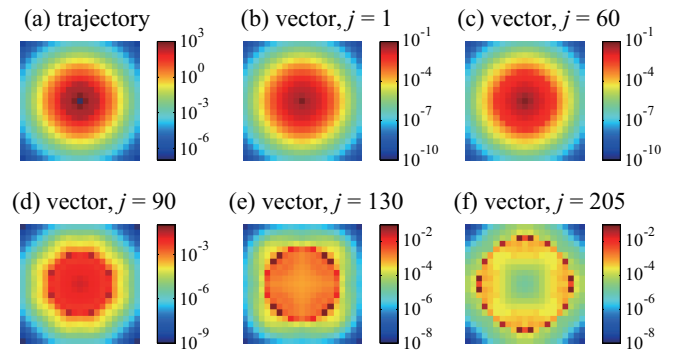}
  \caption{(color online). Power spectrum $S(k_x,k_y)$ of the trajectory (a) and of the CLVs (b-f) for the 2D KS equation. The spectra are shown for $k_x, k_y \in [-k_{\rm cut},k_{\rm cut}]$ with $k_{\rm cut} = 10 \cdot 2\pi/L$.}
 \label{fig:KS2dPowSpec}
\end{figure}%

For $L = 24$ the 2D KS equation shows spatiotemporal chaos
 as in the 1D counterpart studied in Sec.\ \ref{sec:KS1d}.
This leads to a similar structure of the smooth region
 of the Lyapunov spectrum
 [left inset of \figref{fig:KS2dLyapSpec}; cf.\ \figref{fig:KSLyapSpectra}(b)],
 which consists of physical Lyapunov modes carrying information
 on the dynamics.
As for the spurious region,
 despite its apparently intricate structure
 in contrast with the 1D spectrum [see \figref{fig:KSLyapSpectra}(a)],
 we find that the modes therein possess essentially the same
 trivial properties, as shown below.

First, unlike the physical modes which show relatively broad power spectra
 similar to the dynamics [\figref{fig:KS2dPowSpec}(a-c)],
 each spurious mode has a spectrum $S(k_x,k_y)$ sharply peaked
 around a certain wave number $k \equiv \sqrt{k_x^2+k_y^2}$
 [\figref{fig:KS2dPowSpec}(d-f)],
 which is large and increases with the index.
The value of its Lyapunov exponent $\Lambda^{(j)}$ is then
 essentially determined by the linear terms
 of the 2D KS equation \pref{eq:DefinitionKS2d},
 specifically $\Lambda^{(j)} = k^2 - k^4$,
 as indeed confirmed in the right inset of \figref{fig:KS2dLyapSpec}
 (black circles).
Moreover, as in the 1D KS equation, the peak wave numbers $(k_x,k_y)$
 of the spurious modes
 are trivially determined by the geometry:
 for a finite box of size $L_x \times L_y$,
 the wave numbers can take values
 $k_x = 2\pi n_x/L_x$ and $k_y = 2\pi n_y/L_y$
 with intergers $n_x$ and $n_y$.
Their values are then assigned to each mode in such a way that
 the resulting Lyapunov exponent $\Lambda^{(j)} = k^2 - k^4$
 decreases monotonically.
Taking into account the multiplicity for 2D PBC,
 this trivial assignment yields a spectrum indicated
 by the red dashed line in \figref{fig:KS2dLyapSpec},
 which reproduces well the structure of the spurious region.

\begin{figure}[t]
 \includegraphics[width=\hsize,clip]{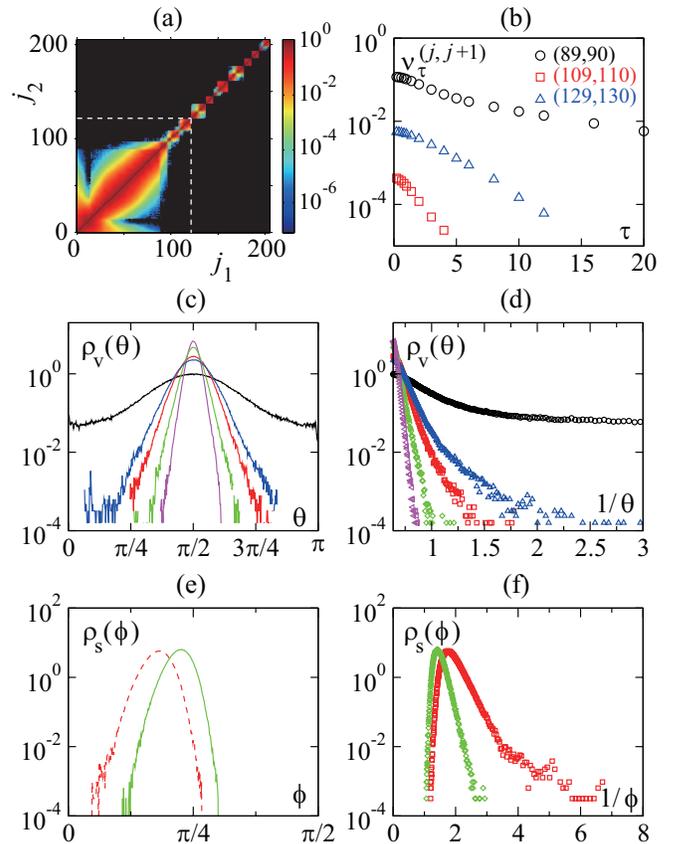}
  \caption{(color online). Hyperbolicity properties for the 2D KS equation. (a) Time fraction $\nu^{(j_1,j_2)}_\tau$ of the DOS violation for arbitrary pairs with $\tau = 0.2$. The black color indicates $\nu^{(j_1,j_2)}_\tau = 0$, i.e., hyperbolically isolated pairs. The white dashed lines indicate the threshold $N_{\rm ph} = 121$ between the physical and spurious modes. (b) Time fraction $\nu_\tau^{(j,j+1)}$ of the DOS violation for pairs of neighboring Lyapunov exponents as a function of $\tau$. (c) Angle distributions $\rho_{\rm v}(\theta)$ for neighboring CLVs of indices $(89, 90), (129, 130), (109, 110), (121,122)$, and $(137, 138)$ from outermost to innermost curve. Notice that they are not in order with respect to the index. (d) $\rho_{\rm v}(\theta)$ against $1/\theta$. The ordinate is averaged over both sides of the distribution. The order of the curves is the same as in (c) from upper right to lower left. (e) Angle distribution $\rho_{\rm s}(\phi)$ between the physical subspace ($1 \leq j \leq N_{\rm ph}$) and the spurious one ($N_{\rm ph}+1 \leq j$) with $N_{\rm ph}=121$ (green solid line). The red dashed line shows the angle distribution $\rho_{\rm s}(\phi)$ between the subspaces of indices $1 \leq j \leq 109$ and $110 \leq j$. (f) The same plots as (e) but shown against $1/\phi$ (green diamonds for the physical and spurious subspaces, and red squares for the other). Each color corresponds to the same index pair throughout in the panels (b-f).}
 \label{fig:KS2dHyperbolicity}
\end{figure}%

To locate exactly the threshold between the physical and spurious modes,
 one needs to measure the hyperbolicity properties
 as in all the other cases.
Figure \ref{fig:KS2dHyperbolicity}(a) shows the time fraction
 $\nu_\tau^{(j_1,j_2)}$ of the DOS violation
 for arbitrary pairs $(j_1,j_2)$.
Although internal connections are found
 for some neighboring steps of spurious modes,
 which are probably caused by mixing of several Fourier modes
 due to accidentally close values of the Lyapunov exponent
 [see \figref{fig:KS2dPowSpec}(d-f)],
 an unambiguous threshold is found at $N_{\rm ph} = 121$,
 for which $\nu_\tau^{(j_1,j_2)} = 0$ for all the pairs
 with $j_1 (j_2) \leq N_{\rm ph} < j_2 (j_1)$
 even at the smallest $\tau$ we used
 [white dashed line in \figref{fig:KS2dHyperbolicity}(a)].
Varying $\tau$ does not unravel
 the internal connections of spurious modes presumably,
 as $\nu_\tau^{(j_1,j_2)}$ decays in the fastest case exponentially
 [\figref{fig:KS2dHyperbolicity}(b)],
 whereas it would decay faster for a hyperbolically isolated pair
 [see Figs.\ \ref{fig:CGLATDosViolation}(a)
 and \ref{fig:CGLPTDosViolation}(a)].

The same conclusion is reached
 from the angle distributions $\rho_{\rm v}(\theta)$
 of the CLVs [\figref{fig:KS2dHyperbolicity}(c,d)].
Pairs of CLVs for physical modes can take any angles and experience
 rather frequent tangencies,
 while the angle distribution for spurious pairs is peaked around $\pi/2$
 and decays quickly in the tail.
A closer look at this tail reveals that hyperbolically isolated pairs,
 or those which fulfill the DOS,
 decay as $\rho_{\rm v}(\theta) \sim \exp(-\const/\theta)$ or faster
 [two innermost plots in \figref{fig:KS2dHyperbolicity}(d)]
 in contrast with the slower decays found
 for pairs of the physical modes (black circles and red squares)
 or the internally connected spurious modes (blue triangles).
The hyperbolic decoupling at $N_{\rm ph} = 121$ is established
 by the same fast decay of the angle distribution $\rho_{\rm s}(\phi)$
 between the two subspaces defined thereby
 [green full curve and diamonds in \figref{fig:KS2dHyperbolicity}(e,f)].
Defining the threshold $N_{\rm ph}$ at a false value leads
 to a clearly slower decay in $\rho_{\rm s}(\phi)$
 [red dashed curve and squares, obtained with the threshold
 corresponding to the red line and squares
 in \figref{fig:KS2dHyperbolicity}(b-d)].
In short, the hyperbolic decoupling between the physical manifold and
 the spurious modes takes place in the same way as for the 1D systems,
 with the peak wave numbers for the spurious modes distributed here
 in 2D Fourier space.

\section{Discussion}  \label{sec:discussion}

In the present paper, we have shown the hyperbolic decoupling
 of the tangent space in spatially-extended dissipative systems
 into a finite set of $N_{\rm ph}$ physical modes
 characterizing all the physically relevant dynamics,
 and a remaining set of spurious modes, which represent
 trivial, exponentially decaying perturbations.
Physical and spurious modes are characterized
 by the existence and absence, respectively, of tangencies between
 the associated CLVs.
In contrast to physical modes which are densely connected to each other
 through frequent tangencies, the spurious modes are hyperbolically isolated
 from all the physical modes, and therefore solely decay to zero
 without activating any physical modes.
The existence or absence of tangencies between Lyapunov modes
 can be investigated
 both from the CLV angle distribution and the violation of DOS.
One can also take into account linear combinations
 of physical or spurious modes
 by measuring the angle between the two subspaces
 spanned by the physical and spurious modes, respectively.
This leads to the unambiguous determination of the number
 of the physical modes, $N_{\rm ph}$.
We conjecture that this number corresponds
 to the minimal dimension of the inertial manifold.
Moreover, we have shown that the hyperbolic decoupling
 between the physical and spurious modes takes place quite generically
 in dissipative systems, as we have found evidence
 in PDEs in one or higher spatial dimensions,
 as well as lattice systems with continuous or discrete time.

A few remarks about related earlier studies are in order.
As already mentioned, the stepwise structure of the Lyapunov spectrum
 in the spurious region was already reported by Garnier and W\'ojcik
 for CGL amplitude turbulence,
 leading them to the conjecture that these are inactive Lyapunov exponents
 \cite{Garnier_Wojcik-PRL2006}.
This is pertinent as we have revealed in the present study,
 but one should be reminded that the stepwise Lyapunov spectrum
 does \textit{not} define or indicate the spurious modes.
The steps do not form at a single well-defined threshold index:
 they are actually slightly inclined, and get flatter and flatter
as one goes deeper into the spurious region.
As a matter of fact, the physical region typically extends to
 the first few steps of the spectrum:
 for the 1D KS system studied here the steps begin at $j=40$ but
 the spurious modes are from $j=44$.
This gap can be quite large; it is found to be $14$
 for the coupled tent maps studied in Sec.~\ref{sec:CML1d}.
The formation of the steps is only a consequence
 of the sharp power spectrum and the choice of the boundary conditions.

In a different work, Ikeda and Matsumoto studied
 the projection of the dynamics onto Fourier modes
 or on the Gram-Schmidt vectors associated with the Lyapunov modes
 \cite{Ikeda_Matsumoto}.
They measured in particular information-theoretic quantities
 and argued that information does not flow from a high wave number region
 to a lower wave number region.
However, similarly to Lyapunov exponents, Gram-Schmidt vectors
 or Fourier modes alone cannot characterize
 the separation between physical and spurious modes,
 and thus cannot provide the exact index of the threshold.
It is the CLVs that bear the true physical properties
 of tangent space and thus are able to define
 unambiguously the physical and spurious Lyapunov modes
 from their hyperbolicity properties.

Concerning the difference between the CLVs and the Gram-Schmidt vectors,
 while it is clear that properties directedly linked to hyperbolicity
 such as angle distributions can only be captured by CLVs,
 one may formally study DOS using either CLVs or Gram-Schmidt vectors.
However, clearly, the finite-time Lyapunov exponents defined
 using the two sets of vectors are \textit{a priori} different
 for any finite times.
Indeed, although these two definitions seem
 to give qualitatively similar results \cite{Kuptsov_Politi-arXiv2011},
 it remains true that the mathematical connection to hyperbolicity
 is only proved for CLVs
 \cite{Pugh_etal-BullAmMathSoc2004,Bochi_Viana-AnnMath2005}.
Thus, in the absence of a detailed study on this point,
 one cannot avoid computing CLVs for drawing any firm conclusion
 on hyperbolicity properties of dynamical systems.

We should also mention that all the angle distributions
 and the finite time Lyapunov exponents used
 for the determination of DOS are not invariant
 against coordinate transformations.
The splitting, however, between physical and spurious modes
 is coordinate independent because by smooth transformations
 a nonzero angle cannot be made to vanish and vice versa.
Furthermore, the type of the essential singularity found
 in the angle distribution of critical pairs
 is invariant against smooth transformations of the coordinates,
 as can be seen by considering
 how the angles are transformed with the coordinates.
A simple example is provided by the two versions of the KS equations,
 \eqsref{eq:DefinitionKS1d} and \pref{eq:DefinitionKS2d},
 which are connected by a linear transformation \cite{Note3}.
These considerations further emphasize the generality of our results. 

Pursuing our conjecture that physical modes form
 the local approximation of the inertial manifold,
 it is crucial to seek a connection between CLVs and phase-space properties,
and in particular to find a way to construct
 the inertial manifold using information of the physical CLVs.
A straightforward construction is a manifold locally defined
 at each point of the trajectory by the physical CLVs,
 but strictly speaking, one also has to consider
 the nonlinear growth of finite-size perturbations,
 which is obviously not captured by Lyapunov analysis.
Because of this,
 it is still in principle possible to consider a system
 in which hyperbolically isolated modes
 correspond to active degrees of freedom,
 such as a lattice of uncoupled hyperbolic maps as a trivial instance.
However, given that the generic systems we studied are by no means hyperbolic,
 it is unlikely that that type of decoupling has any relation 
 to the hyperbolic decoupling studied in the present paper.

Our results are also important from a practical point of view.
Since the physical properties of the PDEs at stake
 are carried by the physical modes,
 a faithful numerical integration needs to incorporate
 at least as many degrees of freedom as the physical modes.
In other words, the number of the physical modes, $N_{\rm ph}$,
 sets a lower bound for the spatial resolution of the simulations.
Moreover, the extensivity of $N_{\rm ph}$ implies that
 measuring it in systems of moderate size
 suffices to have reliable estimates for arbitrarily large system sizes.
Our results may also be useful for controlling spatiotemporal chaos,
 in particular, when determining the minimal number of constraints
 for full control of such a system.
A possible application in this context would be the suppression
 of ventricular fibrillation, the major reason of sudden cardiac death,
 for which promising schemes have been proposed recently
 applying low amplitude electric shocks
 over grids of points or lines \cite{cardiac}.
The number of the physical modes would then help determine
 an appropriate grid resolution.
Note however that, having no access to the basis constituting
 the inertial manifold, one may actually need
 more degrees of freedom or constraints
 than $N_{\rm ph}$ to achieve faithful simulations or full control
 of a dissipative system.
It is therefore of both fundamental and practical importance
 to elucidate the structure of the inertial manifold in phase space
 in the line of the present study.

In summary, we have shown, using Lyapunov analysis and, in particular,
CLVs, that the tangent space of spatially-extended dissipative systems
 can be divided into two parts: a finite-dimensional manifold
 spanned by the physical modes bearing all the physical information
 and the remaining set of spurious, exponentially decaying modes.
We have demonstrated that the spurious modes are hyperbolically
 separated from all the physical modes and hence satisfy DOS.
The separation holds for any linear combination
 of the physical or spurious modes,
 as the minimal angle between the physical and spurious subspaces
 is bounded away from zero.
The number of physical modes has been interpreted
 as the number of effective degrees of freedom needed to describe the dynamics.
This leads to the conjecture that
 the physical modes constitute a local approximation 
 of the inertial manifold.
We hope that these results will trigger mathematical studies
in order to put these issues on a more rigorous basis.
It is also interesting to perform further numerical investigations
 of other dissipative systems such as the Navier-Stokes equation,
 for which no rigorous proof of the existence
 of an inertial manifold is known.

\acknowledgments

The authors thank S. Sasa for informing us of his unpublished work
 on the mode description of solutions of the KS equation
 \cite{Sasa-PC1}.
We acknowledge stimulating discussions
 with P. Cvitanovi\'c, A. Pikovsky, A. Politi, and S. Sasa.

\end{document}